\documentclass[a4paper,11pt]{article}

\usepackage{jinstpub} 

\usepackage{amsmath,amsfonts}
\usepackage{graphicx}
\usepackage{textcomp,nicefrac}
\usepackage{textcomp} 
\usepackage[detect-all,binary-units]{siunitx}
\usepackage{xspace}
\usepackage{arydshln}
\usepackage{notoccite}
\usepackage{booktabs}
\usepackage{array}
\usepackage{adjustbox}
\usepackage{soul}
\usepackage{xcolor}
\usepackage{xspace}

\def\IsDraft{0}
\AtBeginDocument{%
    \if\IsDraft0
        \renewcommand{\st}[1]{\leavevmode\unskip\ignorespaces\xspace}
        
        \renewcommand{\textcolor}[2]{#2}
        
    \else 
        \usepackage{lineno}
        \linenumbers
    \fi
}

\newcolumntype{C}[1]{>{\centering\let\newline\\\arraybackslash\hspace{0pt}}m{#1}}
\title{Timespot1: A \SI{28}{\nano \meter} CMOS Pixel Read-Out ASIC \\ for 4D Tracking at High Rates}

\author[a,1]{Sandro Cadeddu,\note{Corresponding author.}}
\author[b,c]{Luca Frontini,}
\author[a]{Adriano Lai,}
\author[b,c]{Valentino Liberali,}
\author[d]{\\ Lorenzo Piccolo,}
\author[d]{Angelo Rivetti,}
\author[e]{Jafar Shojaii,}
\author[b,c]{Alberto Stabile}
\affiliation[a]{INFN Sezione di Cagliari, 09042 Monserrato, Italy}
\affiliation[b]{Universit\`a degli Studi di Milano, Dipartimento di Fisica, 20133 Milano, Italy}
\affiliation[c]{INFN Sezione di Milano, 20133 Milano, Italy}
\affiliation[d]{INFN Sezione di Torino, 10125 Torino, Italy}
\affiliation[e]{Swinburne University of Technology and University of Melbourne, Victoria, Australia}
\emailAdd{sandro.cadeddu@ca.infn.it}

\newcommand{\rnote}[1]{\textcolor{red}{#1}}
\newcommand{\cor}[2]{\st{#1}\textcolor{red}{#2}}
\DeclareSIUnit\permille{\text{\textperthousand}}

\abstract{We present the first characterization results of Timespot1, an ASIC designed in CMOS \SI{28}{\nano \meter} technology, featuring a \num{32 x 32} pixel matrix with a pitch of \SI{55}{\micro \meter}.
Timespot1 is the first\st{-born} small-size prototype, conceived to \st{read-out} \textcolor{red}{readout} fine-pitch pixels with single-hit time resolution below $\SI{50}{\pico \second}_\text{rms}$  and input rates of several hundreds of kilohertz per pixel. Such experimental conditions will be typical of the next generation of high-luminosity collider experiments, from the LHC run5 and beyond.
Each pixel of the ASIC \st{has been endowed with} \textcolor{red}{includes} a charge amplifier, a discriminator, and a Time-to-Digital Converter with time resolution \rnote{indicatively} \st{around} \textcolor{red}{of} $\SI{22.6}{\pico \second}_\text{rms}$ and maximum \st{read-out} \textcolor{red}{readout} rates (per pixel) of \SI{3}{\mega \hertz}. To respect system-level constraints, the timing performance \st{have}\textcolor{red}{has} been obtained keeping the power budget per pixel below \SI{40}{\micro \watt}.
The ASIC has been tested and characterised in \textcolor{red}{the} laboratory concerning its performance in terms of time resolution, power budget and sustainable rates. 
The ASIC will be hybridized on a matched \SI{32 x 32} pixel sensor matrix and will be tested under laser beam and Minimum Ionizing Particles in the laboratory and at test beams. 
In this paper we present a description of the ASIC operation and the first results obtained from characterization tests concerning its performance \st{in tracking measurements}.
}
\keywords{Front-end electronics for pixel read-out, time resolution, pixel with timing}

\begin{document}
\maketitle
\flushbottom

\section{Introduction}
Vertex detectors of the next generation of collider experiments will have to cope with an increased amount of tracks per event. To solve the problem of event pile-up, trackers must operate with pixel sensors having high space-time resolution. Typical requirements are space resolutions of about \SI{10}{\micro \meter} and time resolutions below $\SI{50}{\pico\second}_\text{rms}$ per hit~\cite{LHCB_TDR}. Furthermore, this level of performance must be reached also at relatively high input hit rates (hundreds of kilohertz), while keeping the power density under control, typically well below \SI{2}{\watt\per\centi\meter\squared}. 

Dedicated development activities have already started to study possible technical solutions in this respect. The TimeSPOT project~\cite{TIMESPOT2018} aims at producing a small-scale tracking demonstrator, consisting \st{in}\textcolor{red}{of} 8 tracking layers, each featuring a \st{sensible}\textcolor{red}{sensitive} area of about \SI{4}{\milli \meter \squared}. Each layer will be based on a matrix of \num{1024} pixels each having a size of \SI{55 x 55}{\micro \meter \squared} and a pixel \st{read-out}\textcolor{red}{readout} chip (the Timespot1) satisfying the performance requirements mentioned above. The characteristics and performance of the TimeSPOT sensors, based on 3D-trench geometry, are widely described in~\cite{LAI2020164491}, \cite{3D-TRENCH}, and~\cite{3D-accurate}. They have shown intrinsic time resolution around $\SI{10}{\pico\second}_\text{rms}$ \cite{3D-accurate}, have a typical pixel capacitance of \SI{100}{\femto\farad} and \textcolor{red}{a Most Probable Value (MPV) signal charge value of \SI{2}{\femto\coulomb}}. 

The Timespot1 scheme consists of a charge pre-amplification stage followed by a discriminator and a TDC for the digitization of the timing information. The electronics should have a performance adequate to keep an overall time resolution well below $\SI{50}{\pico\second}_\text{rms}$ on the full chain, that is including the contribution of sensor, front-end jitter and Time-to-Digital Converter (TDC) resolution. However, due to the very high resolution of the sensor itself, its contribution along the full chain can be considered negligible. Most importantly, the power budget available to the electronics is constrained by the power dissipation system employed in the experiments\st{, corresponding to a power consumption limit around 50uW per pixel}.

\rnote{
A \SI{28}{\nano\meter} CMOS technology was chosen for the ASIC development. Among the nanoscaled technologies, the \SI{28}{\nano\meter} has proven to be the most radiation resistant~\cite{bib:TID_28}. This property is connected to the fact this node is the most scaled one featuring standard planar MOS transistors \cite{bib:65rad1}\cite{bib:65rad2} which performs better when compared to three-dimensional ones~\cite{bib:16rad}. Moreover, this technology was chosen instead of less scaled ones due to the added advantages of increasing the overall transistor density and the digital circuits power efficiency.}

\newcommand{\nota}[1]{\color{blue}\small\textit{#1}\normalsize\color{black}}

\section{State of the art of high time resolution ASICs}

 In the framework of developments on solid-state radiation detectors, several Front-End ASICs \st{have been already} \textcolor{red}{have already been} produced, being able to provide both space and time information on the same detected signal. They are aimed mainly at the future upgrades of tracking detectors, and other applications as well, medical imaging \emph{in primis}. The special topic of the present paper, however, concerns the development of \st{read-out}\textcolor{red}{readout} ASICs capable of concurrent space and time measurements in presence of very high input rates and with high pixel granularity. 
 Concerning the problem of high pile-up at the next high luminosity LHC upgrades, the CMS~\cite{bib:cms} and ATLAS~\cite{bib:atlas} experiments have adopted the solution of so-called timing layers, consisting of one single layer of Minimum-Ionising-Particle (MIP) detectors capable of high time resolution (order $\SI{30}{\pico\second}_\text{rms}$). The timing layer is placed just outside the inner trackers and allows using relatively large pixels or pads, in the order of square millimeters. Such \textcolor{red}{a} solution is not suitable for other upgrade projects like the LHCb Upgrade-II~\cite{bib:FTDRLHCbU2} and the NA62 Giga-Tracker~\cite{bib:na62}, where it is necessary to integrate space-time measurement facilities in all the layers and pixels of the inner tracker, with pixel pitches ranging from \SI{40}{\micro \meter} to \SI{300}{\micro \meter} and time resolutions below $\SI{50}{\pico\second}_\text{rms}$ per single pixel. Such pixel detectors, \st{bearing} \textcolor{red}{having} timing facilities at the level of the inner tracking layers, are often referred to as 4D (in the sense of space-time) trackers.
 
 The first complete set of requirements for an ASIC to be conceived for a 4D tracker has been given by the LHCb collaboration and is reported in Table~\ref{tab:specslhcb}~\cite{bib:FTDRLHCbU2}. The table lists the \textcolor{red}{principal sensors and ASIC} requirements established for two possible configurations (Scenarios A and B) of the Upgrade-II of the LHCb inner tracker (also known as VELO). \textcolor{red}{The innermost radius of the VELO is a key driving parameter for the physics performance and the detector technological requirements, in particular rate capability and radiation hardness. Two scenarios are proposed. In the first scenario, Scenario A, the innermost radius is kept at the current value of \SI{5.1}{\milli \meter} and the sensor layout is the same as for VELO Upgrade I. In the second scenario, Scenario B, the radius is relaxed to \SI{12.5}{\milli \meter} in which case the cluster occupancies match those of VELO Upgrade I.} Details are given in~\cite{bib:FTDRLHCbU2}. Table~\ref{tab:specslhcb} \st{values} clearly indicates that, besides the small pitch and the excellent timing performance, other basic requirements must be satisfied, such as high pixel rate, high data bandwidth per ASIC, radiation resistance, and low power consumption.  This aspect defines the maximum hit rate sustainable by the ASIC corresponding to localized bursts of events and the average hit-rate required to be processed without data loss. These performance levels must be achieved while respecting the specifications of power consumption \cor{posed}{mandated} by cooling and power-delivery systems. \st{This} \textcolor{red}{These} limits translate to a maximum power consumption \st{per unit area} (power budget) ranging from \SI{0.1}{\watt / \centi \meter ^2}~\cite{bib:tofhir2x}~to \SI{1.5}{\watt / \centi \meter ^2}~\cite{bib:FTDRLHCbU2}~\cite{bib:lhcbcooling}. Furthermore, the ASIC architecture and technology must be sufficiently radiation-hard to sustain \st{an} \textcolor{red}{a} total dose in the order of tens of megagrays~\cite{bib:FTDRLHCbU2}.
 
 \begin{table}[tb]
    \centering
    \caption{List of ASIC requirements for LHCb VELO U2} \label{tab:specslhcb}
   \begin{tabular}{lrrl}
     \toprule
        Requirement & Scenario A & Scenario B \\
     \midrule
        Pixel pitch [\si{\micro \meter}]             & $~\leq$\,\num{55}                  & $\leq$\,\num{42}\\
        Matrix size                             & \num{256 x 256}           & \num{355 x 355}\\
        Time resolution RMS [$\si{\pico\second}_\text{rms}$]                & $\leq$\,\num{30}                 & $\leq$\,\num{30}\\
        Loss of hits [\%]                       & $\leq$\,\num{1}                  & $\leq$\,\num{1} \\
        TID lifetime [\si{\mega\gray}]                      & $>$\,\num{24}                    & $>$\,\num{3} \\
        ToT resolution [bits] & \num{6} & \num{8}  \\
        Power budget [\si{\watt / \centi\meter\squared}]                 & \num{1.5}                       & \num{1.5}  \\
        Power per pixel [\si{\micro\watt}]         & \num{23}                        & \num{14} \\
        Threshold level [$\mathrm{e^-}$]                & $\leq$\,\num{500}                & $\leq$\,\num{500} \\
        Pixel rate hottest pixel [\si{\kilo\hertz}]          & $>$\,\num{350}                   & $>$\,\num{40} \\
        Max discharge time [\si{\nano\second}]                 & $<$\,\num{29}                    & $<$\,\num{250} \\
        Bandwidth per ASIC \textcolor{red}{with area} of \SI{2}{\centi\meter\squared} [\si{\giga bit\per\second}]   & $>$\, \num{250}                   & $>$\,\num{94} \\
     \bottomrule
   \end{tabular}
\end{table}
  
Such concurrent requirements give rise to various challenging trade-offs that must be overcome.  The spatial resolution can be improved by reducing the pixel pitch. However, the increase in channel density will affect the power consumption per unit area and produce more local data that will tend to clog up the internal data network.  In turn, a larger pixel will intercept more hits for a given time, increasing the amount of \st{signals} \textcolor{red}{hits} that must be processed by \st{the} \textcolor{red}{a} single channel. Moreover, depending  on sensor type and geometry, the \st{sensible} \textcolor{red}{sensitive} area size will affect signal strength and input impedance which will \st{affect} \textcolor{red}{influence} the Signal-to-Noise-Ratio (SNR) of the very-front-end.  Another challenge is related to improving the time resolution. The two main factors in the design of a high-resolution timing front-end are the input signal SNR and the power budget, which is related to the slew-rate of the amplifier.  Analog circuit SNR and bandwidth, and digital circuit operating frequency are key values for the digitisation of the timing properties of the signal.  As explained earlier, these aspects are tightly related to the channel area and thus to the sustainable hit-rate. 

To reach the desired time resolution, additional circuits for calibration and correction may be required. Such additional features take additional area inside the pixel. High-precision time measurements require to incorporate methods for signal time-walk compensations, such as a constant fraction discriminator or a Time-over-Threshold (ToT) correction. This second method \st{would require} \textcolor{red}{requires} to implement ToT measurement in addition to the Time-of-Arrival (TA) \st{one}.
Moreover, techniques or mitigation of radiation effects, such as triple redundancy in logic and stored configurations, have an additional cost in terms of area.
Overall, the increase in pixel density, complexity and precision will require longer data words that will in turn affect the ASIC throughput.
 
 As of today, no device has been \st{yet} produced nor designed matching such set of requirements. However, since a couple of years, a number of developments have been \st{carried on} \textcolor{red}{made}, including the possibility to measure time with moderate to high resolution. 
 \st{The values range from 100ps to 30ps for the time resolution and down to 10um for spatial resolution.} 
 
\st{The different} \textcolor{red}{Different} R\&D projects focus on different specifications tied to the target experiments or role inside the specific tracking system. High granularity ASICs are \st{characterizes} \textcolor{red}{characterized} by a pixel pitch in the scale of tens of micrometers while low-granularity ones will target resolution\textcolor{red}{s} around one millimeter. The sensor type will also affect the input signal characteristics and input impedance, and define\textcolor{red}{s} the minimum achievable pitch and radiation hardness level. Monolithic \cor{sensors}{devices}, due to the absence of \st{wafer coupling} \textcolor{red}{bump bonding}, exhibit\st{s} a lower \st{input impedance} \textcolor{red}{capacitance} and can be realized in extremely small form factor. The drawback of \st{the} this \cor{approach}{technology} is a relatively low\st{er} radiation hardness~\cite{bib:MonolithicRadHard}, \textcolor{red}{issues of having active electronics on top of on-chip sensor}, \rnote{and a much lower integration capability}. Hybrid \st{sensor enable to couple } \textcolor{red}{pixel detectors couple} a CMOS front-end chip \st{with} \textcolor{red}{to} a sensor matrix built with different processes and technologies. This \st{will enable to connect} \textcolor{red}{enable connecting} sensors made in materials alternative to silicon such as diamond. The shape of the \st{sensible} \textcolor{red}{sensitive} area can also be different from the planar one, such as in 3D sensors~\cite{bib:3d}. The material can also be engineered to add an intrinsic gain layer inside the sensor like in LGAD. For large pitches \textcolor{red}{it} is also possible to employ composite sensors like SiPM and MCP. 

  \newcommand{\tx}[0]{\relax\ifmmode \times \else $\times$\fi}
\newcommand{\ep}[1]{\relax\ifmmode \times 10^{#1} \else $\times 10^{#1}$\fi}
\newcommand{\na}[0]{n.a.\xspace}

\begin{table}
\caption{State of art of Timing Front-End ASICs}
\label{tableState}
\setlength{\tabcolsep}{2pt}

\begin{adjustbox}{max width=\columnwidth}
\begin{tabular}{lcccccc}
\toprule
ASIC name & year & node & \textcolor{red}{rms} time resolution & pixel size & \# of pixels & \textcolor{red}{Time Walk} \st{TW} correction  \\ 
 & & [\SI{}{\nano\meter}]& [\SI{}{\pico\second}] & [\SI{}{\micro \meter}] &     &  \\ 

\midrule

\textbf{Timespot1}$^{ab}$ & 2021 & 28  & \st{50} \textcolor{red}{48}   & \SI{55 x 55}  & 1024 & ToT \\

Timepix4$^{b}$ \cite{bib:tpix_42022} & 2020 & 65  & \st{135} \textcolor{red}{$\sim$125}  & \SI{55 x 55} & \num{229e3}  & ToT \\
FASTPIX$^{c}$ \cite{bib:fastpix} & 2021 & 180  &  142  & 10 to 20 & 68 & ToT \\
FAST2 \cite{bib:fast2}     & 2021 & 110  & 36  & \SI{500 x 500} & 32          & soft CFD \\
FastIC$^{c}$ \cite{bib:fastic} & 2021 & 65 & 107 & \SI{1500 x 1500} & 8 & ToT \\
DIAMASIC \cite{bib:diamasic} & 2021  & 130 & 80  & \na & \na & no \\
TOFHIR2X \cite{bib:tofhir2x} & 2021  & 130   & 24$^{d}$, 55$^{e}$  & \SI{3000 x 3000} & 32 & amplitude \\
ETROC1$^{c}$ \cite{bib:etroc1} & 2021 & 65  & 29  & \SI{1300 x 1300} & 16 & ToT \\
ALTIROC1 \cite{bib:altiroc1} & 2020 & 130  &  35$^{d}$, 70$^{e}$ & \SI{1300 x 1300} & 25 & ToT \\
FCFD0$^{bc}$ \cite{bib:fcfd0} & 2021 & 65 &  30 & \na  & 3 & CFD  \\
\textcolor{red}{TDCpix \cite{TDCpix}}& 2013 & 130 & $\sim$200  & \SI{300 x 300} & 1800 & ToT\\
\bottomrule

\multicolumn{7}{p{\columnwidth}}{The ASICs presented in this table feature a measured time resolution better than \SI{200}{\pico\second}. Measurements have been performed with an actual sensor and radiation source or with an electrical auto-test. Data are flagged as \na when either not available or not applicable.}\\
\multicolumn{7}{p{\columnwidth}}{$^{a}$This work, $^{b}$Electrical auto-test only, $^{c}$no TDC, \textcolor{red}{$^{d}$Beginning of Life (zero dose) , $^{e}$End of Life (maximum dose).}}
\end{tabular}
\end{adjustbox}
\label{tab:review}
\end{table}

 Table \ref{tab:review} summarises the state of the art of \st{the} timing front-end ASICs already developed and produced, featuring a measured time resolution better than \SI{200}{\pico\second}. \st{The measured results concerning their main characteristics are reported.}
 Among such developments, Timespot1 is a prototype ASIC for high-granularity high-hit-rate pixel-matrix, specifically developed for the readout of a high time resolution matrix of 3D-trench silicon pixels. The usage of a 3D silicon sensor opens up the possibility to use this chip in a high radiation environment. In terms of time resolution the proposed pixel architecture is able to gain at least a factor two when compared to other similar ASICs of the same category. \st{Overall its timing performance indicates that the proposed ASIC could be a good candidate as a general purpose 4D-pixel front-end} \textcolor{red}{Its timing performance indicates that the proposed ASIC will be a good candidate for future 4D-pixel detectors}.

\section{Chip architecture}

\begin{figure}
    \centering
    \includegraphics[width=0.7\columnwidth]{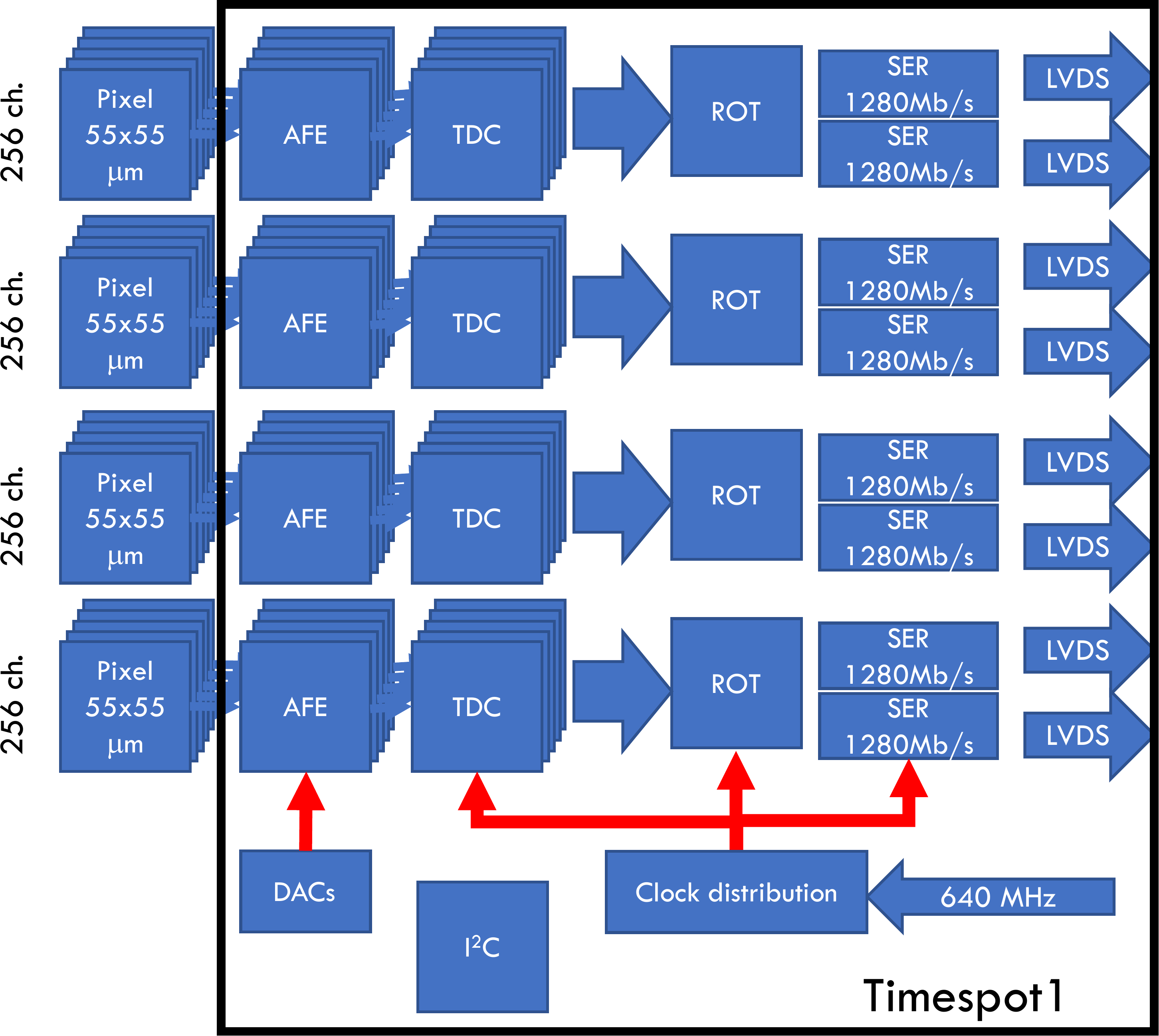}
    \caption{Chip block architecture, The input channels are organized in 4 groups of 256 channels. Each group is connected to its own Read Out Tree (ROT), which sends \textcolor{red}{the collected data} to the serializers (SER) \st{the data collected}.}
    \label{fig:Timespot1BlkArch}
\end{figure}

\begin{figure}
    \centering
    \includegraphics[width=1\columnwidth]{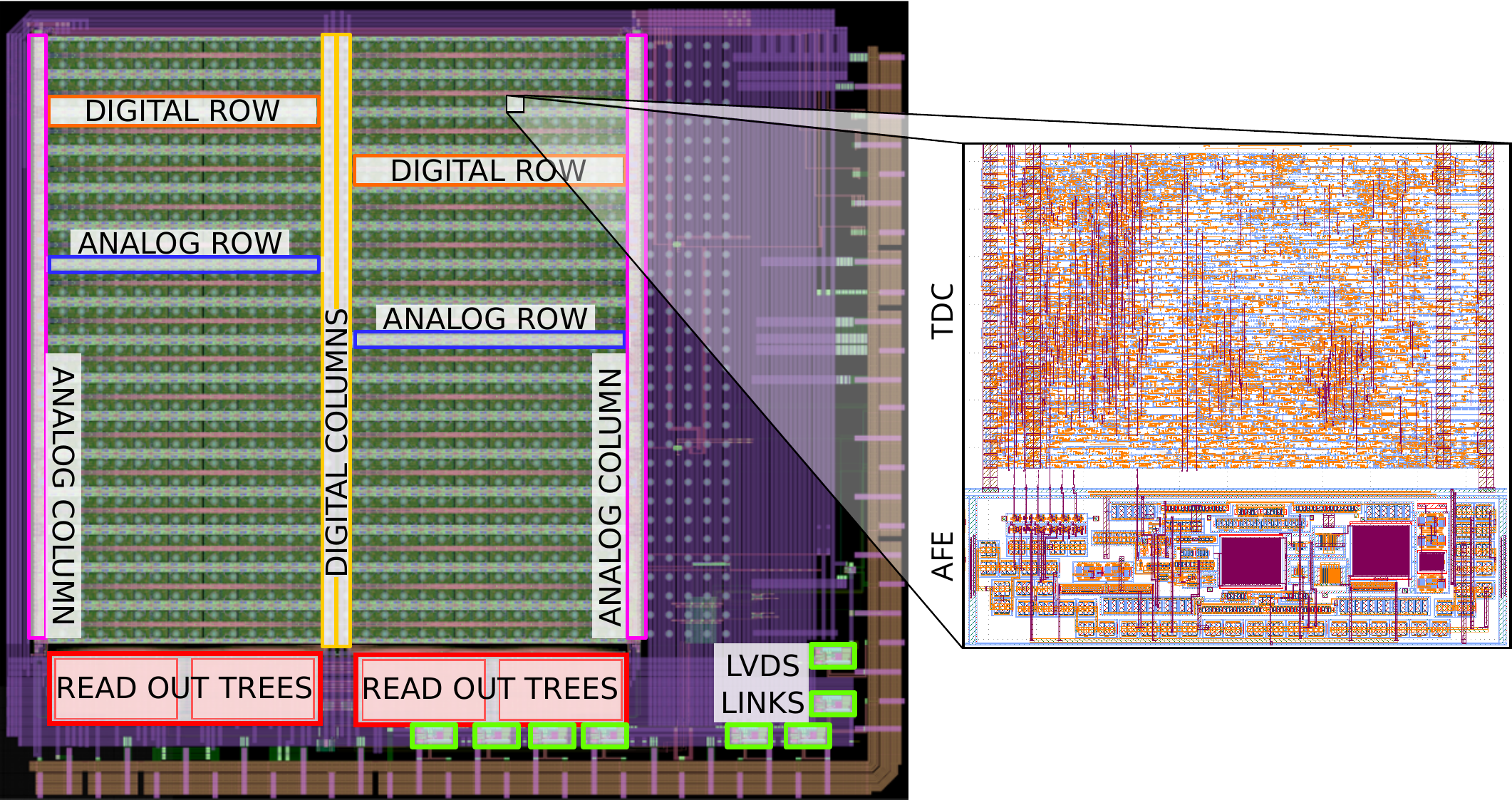}
        \caption{Full chip layout, with the main structures highlighted \textcolor{red}{and the layout of the pixel cell}. \st{It} \textcolor{red}{The chip} \st{sizes} \textcolor{red}{has dimension of} \SI{2.618 x 2.288 }{\milli \meter \squared}.}
    \label{fig:labeled_chip}
\end{figure}

The Timespot1 ASIC is designed to provide a readout array\st{,} suitable for chip-to-chip bump-bonding with 3D pixel arrays\st{,} with a pixel size of \SI{55 x 55}{\micro\meter\squared}. \st{The Timespot1 chip} \textcolor{red}{It} integrates \num{1024} channels, organized in a \num{32 x 32} matrix, with each channel equipped with its own Analog Front-End (AFE) and TDC. The block architecture is shown in Figure \ref{fig:Timespot1BlkArch}, while Figure~\ref{fig:labeled_chip} shows the full layout of the chip, featuring \num{9} metal levels.

The input channels are organized in two blocks of 512 pixels, each \st{one} consisting of 2 groups of 256 channels. Each group is connected to one out of four Read Out Tree (ROT) blocks. Each ROT collects data from the active channels, assigns them a global timestamp and sends formatted data to one of the two serializers connected to LVDS drivers\st{, so to output data towards the acquisition system}. In total, 8 LVDS drivers are integrated, sustaining a data rate of \SI{1.28}{\giga \bit \per \second} each, with an overall \st{nominal} \textcolor{red}{maximum} data throughput of \SI{10.24}{\giga \bit \per \second}. 

From the floorplan point of view, the \num{32 x 32} array is arranged in two mirrored \num{16 x 32} pixels blocks. The size of each channel is \SI{50 x 55}{\micro\meter\squared}, therefore the pitch is reduced in the horizontal direction compared to the bond-pad matrix. A redistribution layer is used to connect each channel to its own bond-pad.  In this way, \SI{80}{\micro\meter} are saved every 16 pixels in order to be used for distribution of analog references and power supplies on one side (Analog Column), and for digital power supplies and data transmission lines on the other side (Digital Column). Particular care was used to maintain as separate as possible the digital domain from the analog one. In particular, the analog part of the channels was integrated in an Analog Row with its own substrate bias and power nets\textcolor{red}{, connected only with the Analog Column}. In the same way the digital part (TDC and control logic) is confined in its dedicated area (Digital Row)\textcolor{red}{, which is connected only with the Digital Column}. \st{The interconnections to the respective service columns were realized only in the opposite sites, making them independent.} \textcolor{red}{In this way Analog and Digital domains are kept independent.} 

\rnote{
Particular attention was paid in regard to the coupling between the sensor and the pre-amplifier, since the redistribution scheme would create differences in the interconnection paths.
The individual interconnection wires have been optimized in order to control their parasitic capacitance.
 The total capacitance due to the sensor together with the bond pad is below \SI{150} {\femto\farad}, therefore the small capacitance differences (of the order of a few femto-Farad) due to the differences in the interconnections, can be neglected. }

The readout scheme is totally data-driven and trigger-less, with time-stamping from an externally provided signal.
Timespot1 also integrates \st{several} structures for internal services\st{:}\rnote{, such as} two voltage bandgap\textcolor{red}{s}\cor{,}{ and} 8 DACs for Voltage references\st{, one PLL and two DCO generating the needed clock frequencies}.

The chip configuration is handled using a slow-control interface I\textsuperscript{2}C-like protocol~\cite{NXPI2C7}.
\st{The top level and each Digital Column and Row features its own independent target interface with its unique address.}

\subsection{Analog Front-End}

\begin{figure}
    \centering
    \includegraphics[width=0.9\columnwidth]{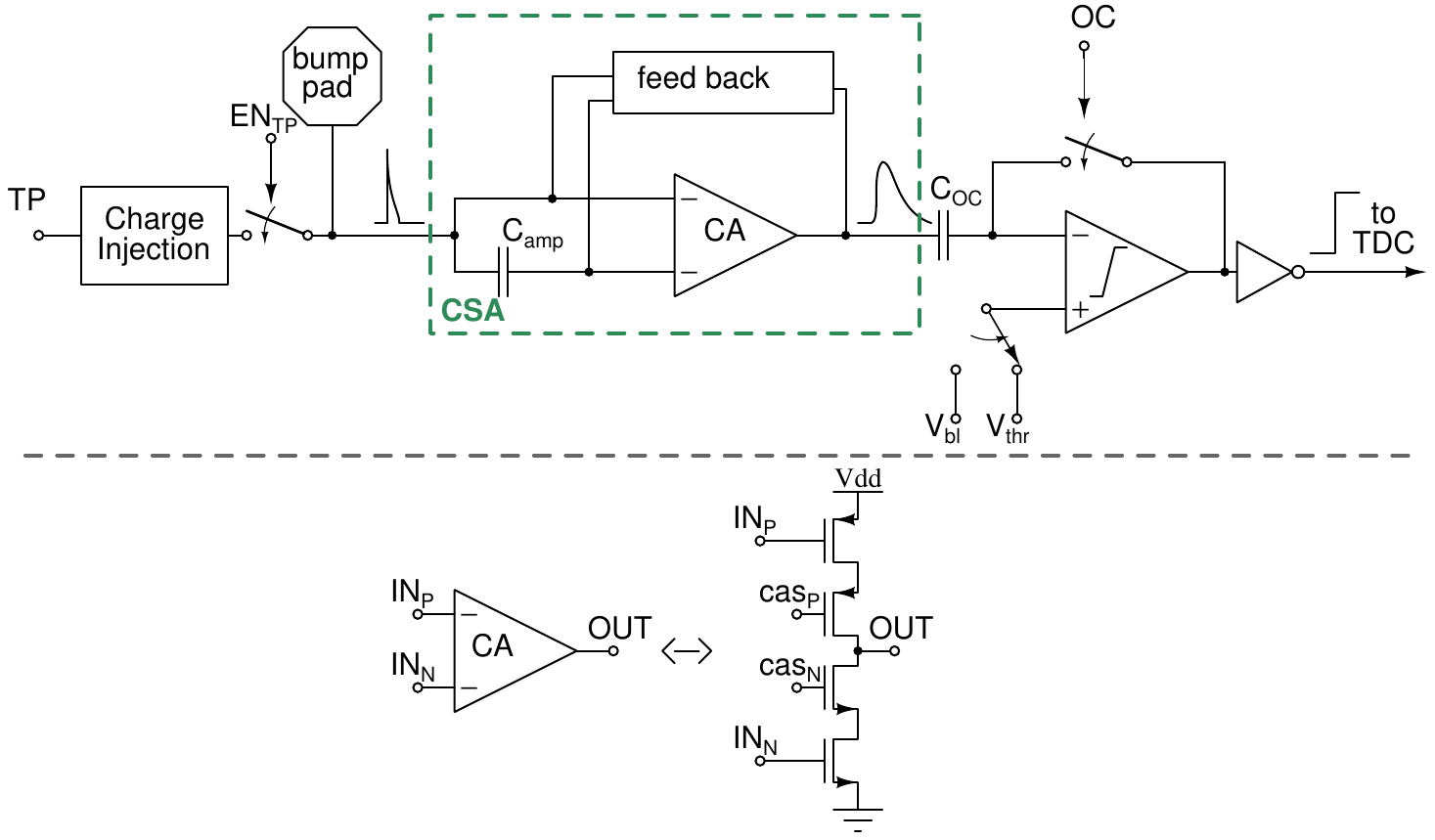}
        \caption{Schematic block representation of the AFE. \rnote{Each channel can be powered off individually.    The block labeled as CA is the Core Amplifier of the pre-amplifier, its transistor level schematic is shown below.  The $EN_{TP}$ signal is used for connecting the charge injection circuit.  TP is a digital signal used to trigger the generation of the current test pulse.    The OC signal is used to initiate the offset compensation procedure.}}
    \label{fig:afe_sch}
\end{figure}

A schematic representation of the \st{AFE} \textcolor{red}{Analog Front-End (AFE)}  \cite{bib:phdlpiccolo} is shown in Figure~\ref{fig:afe_sch}. Each channel is made of a Charge Sensitive Amplifier (CSA) as first stage and a leading edge discriminator as second stage. Additionally, each CSA is both connected to the respective sensor bond-pad and to its own charge-injection circuit. This last block is used to generate current test pulses to be used during electrical self-tests. The total area is \SI{15x50}{\micro\meter^2} which also includes the required digital buffers and decoupling capacitors. 

The CSA is a trans-impedance stage which produces output voltage signals proportional to the integrated charge of the input current signals.
\st{The proportionality factor is the inverse of the feedback capacitance connected between the CSA input and output nodes.} \cor{The output SNR is proportional to this gain factor, whereas the output jitters is inversely proportional. The  charge-to-voltage gain can be maximized by minimizing the feedback capacitance, therefore no additional capacitors is inserted in the feedback.  The capacitive feedback path is formed by the input transistors gate-drain and parasitic capacitance.}{In terms of jitter performance a key parameter is the total capacitance coupling the input and output nodes of the CSA.   The direct effect is connected to the fact that the output signal slew-rate is inversely proportional to this capacitance. Therefore, this total capacitance must be minimized. For this reason, no additional capacitor was added in feedback, leaving only the parasitic capacitance coupling the two nodes.} The capacitance value is estimated \rnote{using a parasitic extraction method} to be around \SI{3}{\femto \farad}.
The CSA consists in a Core Amplifier (CA) and a Krummenacher filter \cite{bib:krum}. 

The function of the CA is to charge the feedback capacitance. \cor{In order to produce signals with high time accuracy, this block features high gain, high band-width and low noise.}{The output noise and bandwidth of this block are critical to enhance the signal time accuracy.   In fact, the time jitter of the pre-amplifier stage can be quantified by the ratio between its signal noise and slew-rate.   }%
These characteristics are all tied to the available power, therefore it is critical to maximize the architecture performance for a power consumption under a constrained power budget of \SI{15}{\micro \watt}.
\rnote{Moreover, in the CSA configuration, the core amplifier open loop gain masks the stage input capacitance which makes possible to obtain higher slew-rates with the same power.}
The block has been implemented using a single inverter-like stage.  Compared to a single input transistor stage, this architectures provides double the transconductance with the same bias current.    In order to boost the open loop gain of the architecture, the two transistors have been cascoded.    A challenging aspect of a linear inverter-based amplifier is the proper biasing of its DC operating point.    In particular, it is difficult to force a reliable DC input voltage while also forcing the desired bias current in the stage. To achieve this, the NMOS input has been AC coupled to the PMOS \st{one}, splitting the DC voltages of the two transistors.  \st{This} \textcolor{red}{These} two nodes are then biased using two separate feedback paths.

The two feedback circuits are implemented using two complementary realizations of the Krummenacher filter in order to have the best DC matching.    The NMOS variant is used to bias the PMOS input, while the PMOS one is used for the NMOS input. The feedback paths are also responsible for the discharge of the feedback capacitor.   In particular, this architecture features a constant current discharge which creates a proportionality between the input charge and the output signal ToT. The discharge current value is programmable and can be set between \SI{25}{\nano\ampere} and \SI{100}{\nano\ampere}. Additionally, this architecture is also able to compensate the sensor leakage current up to values around to its discharge current\rnote{: up to \SI{100}{\nano\ampere}}.

The \st{leading edge} discriminator produces the digital pulses which are used as the TDC input.  These pulses must retain the timing properties of the analog signals.   The CSA signal is compared to a given voltage threshold, producing a \cor{ high slew-rate} { steep digital step} when the input exceed the threshold. The core of this block is implemented with a two stage open loop amplifier with differential input.  The first differential stage is used to compare the signal and the threshold. The second single ended stage is used to boost the voltage gain \rnote{bringing the total gain of the discriminator core to \SI{100}{\dB}}.

In order to produce a reliable discrimination across the channels, the effective threshold variation must be accounted for.  This is usually carried out with a per-pixel threshold calibration. Due to the required precision and available pixel area this approach is unfeasible for this application.    The threshold offset was corrected dynamically with a discrete-time offset-compensation circuit.  The Offset Compensation (OC) procedure allows equalizing the CSA output DC levels, and compensating the discriminator offset. This feature is implemented by AC-coupling the CSA output with the discriminator input using the capacitor $C_{oc}$ (Figure~\ref{fig:afe_sch}), and by inserting a switching circuit between the \cor{non-inverting input of the discriminator and its output}{signal terminal and the output of the inverting amplifications stage, thus creating a negative feedback loop}.   During the OC the discriminator is closed in an almost-unitary loop. In this condition the voltage \cor{applied to its}{developing at its} \cor{inverting}{threshold} terminal is saved in the capacitor $C_{oc}$. This voltage will act as the signal baseline level. By reopening the feedback the discriminator is again able to process the incoming signals.  The actual threshold can now be applied to the \cor{inverting}{threshold} terminal: its value will be relative to the saved baseline. From a system-level point of view, this procedure requires only a  common voltage level shared among the channels.  Each channel must only include its own digital control signals.

The baseline value will tend to drift away from the stored one due to the presence of parasitic currents, discharging $C_{oc}$.  Therefore the OC procedure must be repeated periodically. The frequency of this operation and the time required to perform it will determine a dead time for the channel.   \cor{Therefore the switching circuit has been optimized to reduce the dead-time.}{$C_{oc}$ and the feedback switch have been sized in order to optimize the circuit dead time.  Specifically, the sizing of $C_{oc}$ constitutes a trade-off between the the OC coherence time and a low pass filtering by the mean of its parasitic capacitance to ground.   Its capacitance value has been set to \SI{100}{\femto\farad}. The switch has been sized acting on its $W/L$ in order to reduce it off-current and therefore its on-current.  This creates a trade-off between the OC setting time and its coherence. As result the circuit is able to achieve a \SI{100}{\nano\second} setting time, with a coherence time of \SI{1}{\milli \second} corresponding to a dead time of \SI{0.018}{\permille}. This result was measured on a previous prototype \cite{ocmes}.}  

Finally, after the core discriminator, the signal is buffered with a CMOS inverter in order to produce a digital pulse.

\subsection{TDC}

\begin{figure}[t]
    \centering
    \includegraphics[width=0.5\columnwidth]{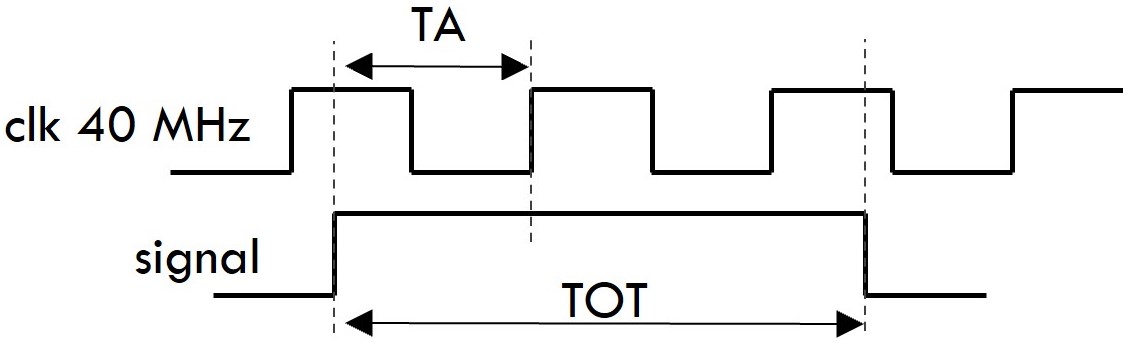}
    \caption{TDC measures signal TA and ToT at the same time.}
    \label{fig:tdcMeasScheme}
\end{figure}
\begin{figure}
    \centering
    \includegraphics[width=0.9\columnwidth]{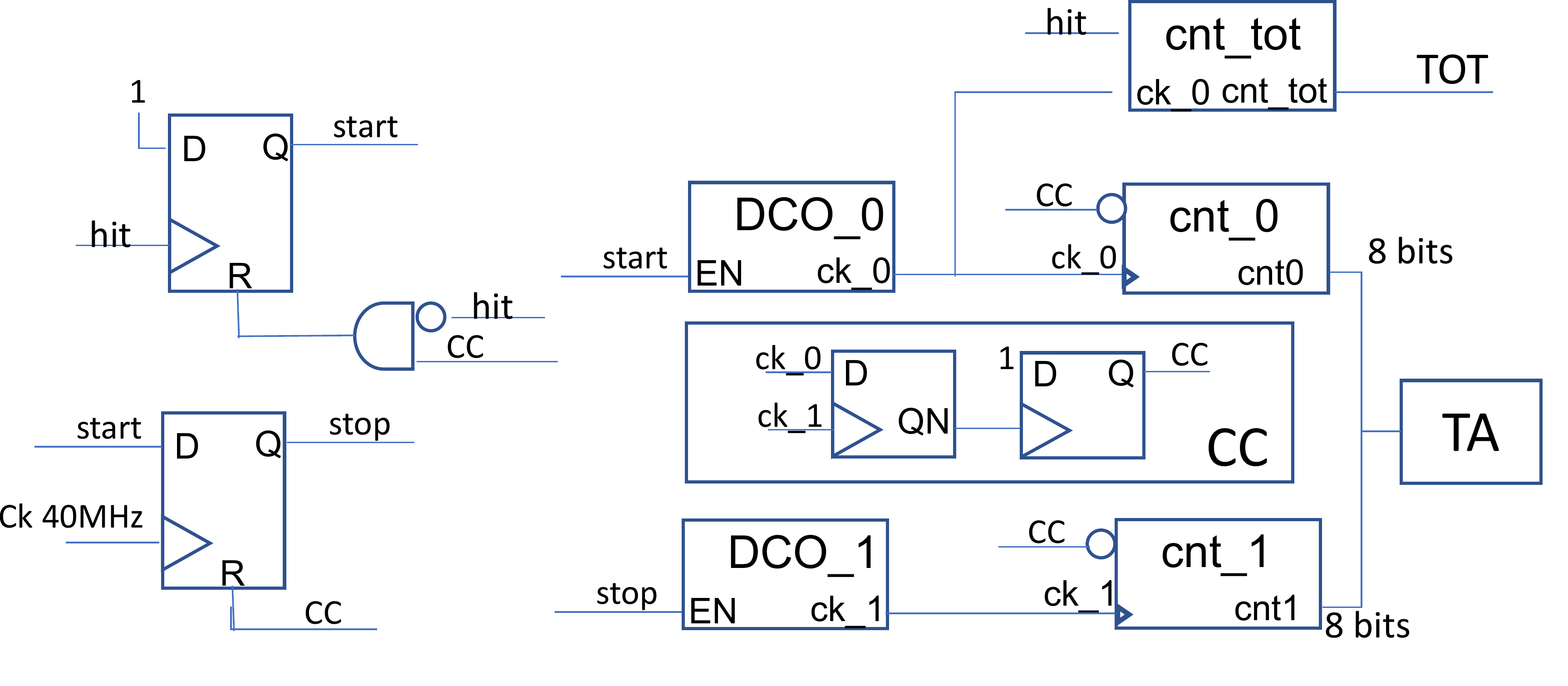}
    \caption{TDC scheme architecture.}
    \label{fig:tdcScheme}
\end{figure}
\begin{figure}[t]
    \centering
    \includegraphics[width=0.65\columnwidth]{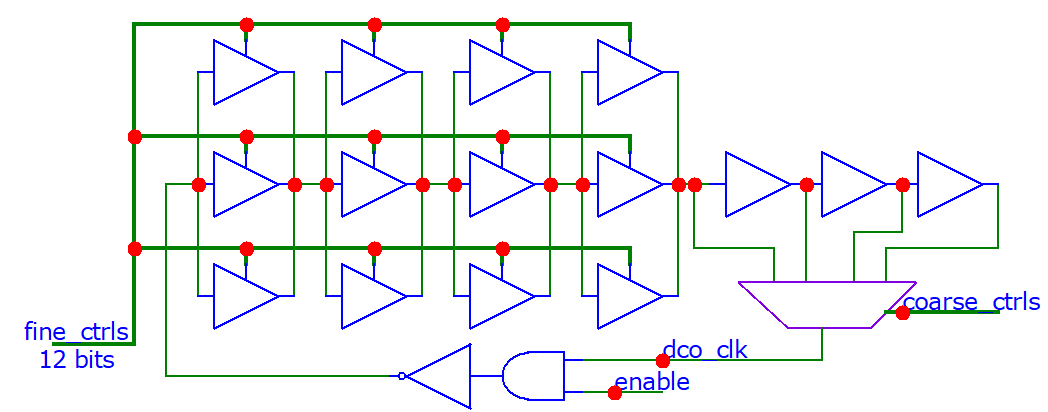}
    \caption{DCO scheme.}
    \label{fig:dcoScheme}
\end{figure}

The TDC measures the Time of Arrival (TA) of the AFE signal, in terms of its phase with respect to a reference clock running at 40 MHz. At the same time the TDC measures the Time Over Threshold (ToT) of the signal to correct the signal Time Walk to improve the TA measurement (Figure~\ref{fig:tdcMeasScheme}). \rnote{The ToT resolution will define the granularity of the TA correction, which in turn will affect the overall detector resolution. The minimum required resolution for the ToT has been evaluated based on the expected AFE TA versus ToT relation to be \SI{0.6}{\nano\second} rms.}

The TDC is based on a Vernier architecture \cite{bib:VernTech}, with two identical Digital Controlled Oscillators (DCOs) working at slightly different frequency (DCO\_1 is faster than DCO\_0) (Figure~\ref{fig:tdcScheme}). The DCOs are made with a tapped delay line with tunable length (Figure \ref{fig:dcoScheme}). Each \st{one} of the first four DCO step element\textcolor{red}{s} is made with three different tri-state buffer\textcolor{red}{s} in parallel, each one with different drive strength. The fine delay is tuned by enabling one \textcolor{red}{of} them at a time.
The last three stages are fixed delay cells and they are used for coarse adjustment. Each DCO drives an \textcolor{red}{8~bit} fast counter. When in stand-by, both DCO\textcolor{red}{'s} are stopped, saving power. The signal edge starts the first DCO, whereas the second is started by the next reference clock edge. A Coincidence detect\st{or} Circuit~(CC) is  used to detect when the two positive edges rise at the same time, indicating the end of measurement. A third counter is driven by the first DCO to calculate the signal ToT.

The TA is computed using:
\begin{equation}    \label{TAmeasFormula}
    TA = (cnt_{0} - 1)T_{0} - (cnt_{1} - 1)T_{1}
\end{equation}
\noindent where \(T_{0}\) and \(T_{1}\) are the periods of the two DCO\textcolor{red}{'s} while \(cnt_{0}\) and \(cnt_{1}\) are the related counter values.
The theoretical resolution of a Vernier TDC is given by the period\st{s} difference:
\begin{equation}
    R_{th} = T_{0} - T_{1}
    \label{vernierResFormula}
\end{equation}
\textcolor{red}{The conversion time for a Vernier architecture is not fixed and can be different for each measure. The maximum conversion time is given by:
\begin{equation}
    T_{max\_conv} = \frac{T_{0}T_{1}}{R_{th}}
    \label{vernierMaxConvTime}
\end{equation}
\noindent where \(R_{th}\) is given by the (\ref{vernierResFormula}).}

\st{The} DCOs are calibrated at the beginning\textcolor{red}{, independently for each pixel,} to set the resolution and all the parameters needed to calculate the time. The calibration process \textcolor{red}{is not aimed to fix precisely the DCOs frequencies, i.e. using some reference signal, but it} sets the period difference ($T_{0} - T_{1}$) between the two DCOs to be consistent with the required resolution. \textcolor{red}{Fixing the difference, instead of a given frequency, makes the system more stable with respect to the environment dependent variables like temperature.}  \st{During calibration the DCO period is measured and adjusted to obtain the desired period difference}. At the end of calibration process, the working periods of the two DCOs (\(T_{0}\) and \(T_{1}\)) are stored \textcolor{red}{on registers inside each pixel} for TA calculation according to (\ref{TAmeasFormula}).
The full calibration\textcolor{red}{, completely managed internally to the pixel,}  needs less than \SI{4}{\micro\second} to be completed and it is performed in two steps: 
\begin{enumerate}
    \item The DCO\_0 (the slowest) is set to generate a clock period \textcolor{red}{around \SI{1000}{\pico\second}. The logic sets the DCO\_0 with the slowest settings and activates it calculating the period. If the period is much higher than \SI{1000}{\pico\second}, the logic act on the coarse setting (Figure \ref{fig:dcoScheme}), reducing the length and consequently the period, and performing a new try.}
    \item The DCO\_1 (the fastest) is set to have a period slightly smaller than the DCO\_0, according to the required resolution. 
    The period is set iteratively while monitoring the period difference $T_{0} - T_{1}$ until the desired one is met. \textcolor{red}{The $T_{0} - T_{1}$ used for the results on this paper is fixed around \SI{50}{\pico\second}, which therefore corresponds to a theoretical LSB of \SI{50}{\pico\second}.}
\end{enumerate}

\rnote{During testing the periods of the DCOs obtained from the calibration showed good stability. By repeating the calibration over time, the variation of the periods always remained in the order of a few ps.}

The TA value\textcolor{red}{, expressed in \si{\pico\second},}  is calculated internally according to (\ref{TAmeasFormula}). The ToT is simply measured by counting the number of DCO\_0 oscillations \st{occurred} \textcolor{red}{occurring} while the input pulse is active.
\textcolor{red}{Both measurements require a time that cannot be foreseen \emph{a priori}. With DCO period values around \SI{1}{\nano\second} and a \(R_{th}\) in the order of \SI{10}{\pico\second}, using the \ref{vernierMaxConvTime} we obtain a \(T_{max\_conv}\) of \SI{100}{\nano\second}. The maximum ToT expected is \SI{200}{\nano\second}. To avoid having to add a pixel-level timestamp, consequently increasing the data to be transmitted, the TDC works with a fixed latency, and the data is published and transmitted after a fixed time, large enough to guarantee the completion of the measurements, both for TA and ToT. The fixed latency determines the dead time of the TDC, and consequently the TDC maximum sustainable rate, which turns out to be \SI{3}{\mega\hertz}.
}

\st{The TDC output word consists of 23 bits where 15 bits are reserved for the TA measurement and 8 bits for the ToT one.}%
\textcolor{red}{The TDC data output includes both TA and ToT measurements. The TA measurement is expressed in ps and, having to cover a dynamic range of \SI{25}{\nano\second}, requires a \SI{15}{\bit} word. The ToT is calculated in terms of the number of cycles of the DCO\_0 clock, whose period is fixed at about \SI{1}{\nano\second} during the calibration phase. Having to cover a range up to \SI{200}{\nano\second}, it requires an \SI{8}{\bit} word. TDC data are output serialized at \SI{160}{\mega\bit\per\second}.}

For debug purpose it is possible to send out the counter values (\(cnt_{0}\) and \(cnt_{1}\)) instead of the TA calculated internally.

\begin{figure}[t]
    \centering
    \includegraphics[width=1\columnwidth]{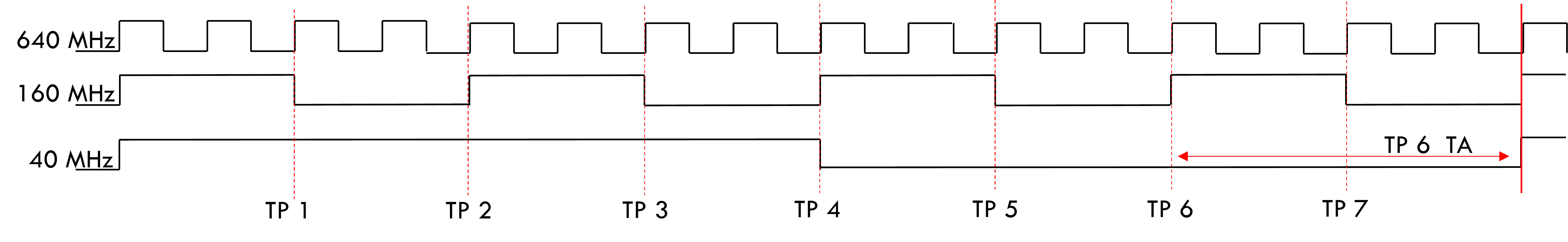}
        \caption{Internal clocks \SI{160}{\mega\hertz} and \SI{40}{\mega\hertz} are derived from the \SI{640}{\mega\hertz} system clock by division. The seven Test Point (TP) are generated on both \SI{160}{\mega\hertz} clock edge, giving a well known Time Interval to be tested, equally distributed over the \SI{25}{\nano\second} dynamic range.}
    \label{fig:TAtestPointGen}
\end{figure}

Furthermore, the TDC is able to self-generate test pulses to verify the TA measurement (with 7 different phases) and ToT measurement (with 32 different widths). 
\textcolor{red}{The seven Test Point (TP) are generated on both \SI{160}{\mega\hertz} clock edges (Figure \ref{fig:TAtestPointGen}), giving a well known Time Interval to be tested, equally distributed in step of \SI{3.125}{\nano\second} over the \SI{25}{\nano\second} dynamic range. Using these 7 points we can test the TDC to get an indication of its linearity and any INL problem. We can also study the distribution of repeated measures by calculating the standard deviation for each test point.}

Table \ref{tab:tdcCharacteristic} summarizes \st{some} \textcolor{red}{the} TDC characteristics and performance. Resolution and power consumption are related to post-layout simulations. \textcolor{red}{Although in the IDLE condition the DCOs are switched off, the different service structures, such as the configuration and the services for the AFE circuit are active and contribute to the consumption. Some of these structures can be shared between different TDCs, thus contributing to a further reduction in overall consumption. This will be implemented in a following release.}
\begin{table}[t]
    \centering
    \caption{Summary of characteristics and performance for the TDC block. Resolution and power consumption are related to post-layout simulations.}
    \label{tab:tdcCharacteristic}
    \begin{tabular}{l c}
        \toprule
        Size &  \SI{50 x 20}{\micro\meter\squared} \\
        Max Event Rate & \SI{3}{\mega\hertz} \\
        Output Word & 24bits Serial @ \SI{160}{\mega\hertz} \\
        TA best \st{resolution} \textcolor{red}{bin size} & \SI{7}{\pico\second} \\
        \textcolor{red}{TA bin size used for prototype test} & $\sim$ \SI{50}{\pico\second} \\
        ToT resolution & $\sim$ \SI{1}{\nano\second} \\
        PWR in IDLE & \SI{20}{\micro\watt} \\
        PWR @ \SI{100}{\kilo\hertz} Event Rate & \SI{26}{\micro\watt} \\
        \bottomrule
    \end{tabular}
\end{table}
\subsection{Readout processing logic} 
\begin{figure}[t]
    \centering
    \includegraphics[width=0.9\columnwidth]{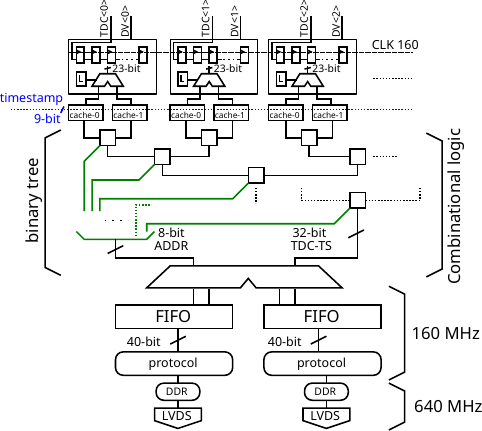}
        \caption{Block diagram of the readout processing logic composed by the ROT and the data serialization circuit.}
    \label{fig:rotScheme}
\end{figure}

Each TDC is connected throughout the Digital Column via its dedicated \textcolor{red}{\SI{160}{\mega\bit\per\second}} serial line \textcolor{red}{and its corresponding Data Valid (DV) signal} to the readout processing logic. At every hit, the TDC sends its \SI{23}{\bit} word, and activates the DV signal.
Figure~\ref{fig:rotScheme} shows the block diagram of the readout processing logic that contains the \st{ROT} \textcolor{red}{binary tree} and the data serialization circuit. \rnote{This block receives data from \num{256} TDCs.}

Each hit-data coming from the TDC is paired with the corresponding timestamp information (\SI{9}{\bit}) that indicates the sequential number of a counter on the \SI{40}{\mega\hertz} clock when the hit data is generated. This counter \cor{can be}{is} started using a dedicated external signal.  \textcolor{red}{The timestamp information is added} \st{This step is required} at this point since the successive logic does not have a fixed latency.
The data is then cached in order to reduce the risk of data-loss.
While data are stored into the cache memories, a\st{n asynchronous} \st{ROT}\textcolor{red}{combinational binary tree}~\cite{FISCHER2001499} sequentially reads the data from the \st{buffers} \textcolor{red}{caches} at a rate of \SI{160}{\mega\hertz} and frees them.
The \st{ROT}\textcolor{red}{binary tree} generates the geographical coordinates of the pixel (\SI{8}{\bit}) and implements a zero-suppression feature.

The output of the  \st{ROT}\textcolor{red}{binary tree} consist of: \SI{8}{\bit} address, \SI{23}{\bit} TDC data and \SI{9}{\bit} timestamp information. The resulting \SI{40}{\bit} string is fed to one of two \num{32}-word-deep FIFO\textcolor{red}{'s}, working at \SI{160}{\mega\hertz}. 
The FIFOs are used to mitigate activity peaks that otherwise could cause data loss.
The \SI{40}{\bit} output of the FIFO is encoded using a custom transmitting protocol. 
The protocol is organized in \SI{8}{\bit} words and it is constructed as follows: when data is present at the FIFO output, it is divided in five bytes preceded by a header byte, otherwise an idle byte is transmitted. 
The idle and header words can be chosen and set using the slow control.

The protocol block outputs a new byte at a frequency of \SI{160}{\mega\hertz}. Each byte is then serialized with a \SI{640}{\mega\hertz} Double Data Rate (DDR) and transmitted to the output using an LVDS protocol at \SI{1280}{\mega\bit\per\second}.
The maximum output bandwidth of Timespot1 is defined by the 8 LVDS cumulative bandwidth, resulting in \SI{10.24}{\giga\bit\per\second}. 
The per-channel sustainable hit-rate by the read out processing logic ranges from \SI{3}{\mega\hertz} to \SI{200}{\kilo\hertz} depending on the channel occupancy. The first \st{case} \textcolor{red}{value} corresponds to a single active pixel whereas the second ones correspond to a totally uniform hit occupancy (all channels firing at the same time).

\section{Chip Measurements}

\subsection{Setup and Methods}

\begin{figure}[t]
    \centering
    \includegraphics[width=0.7\columnwidth]{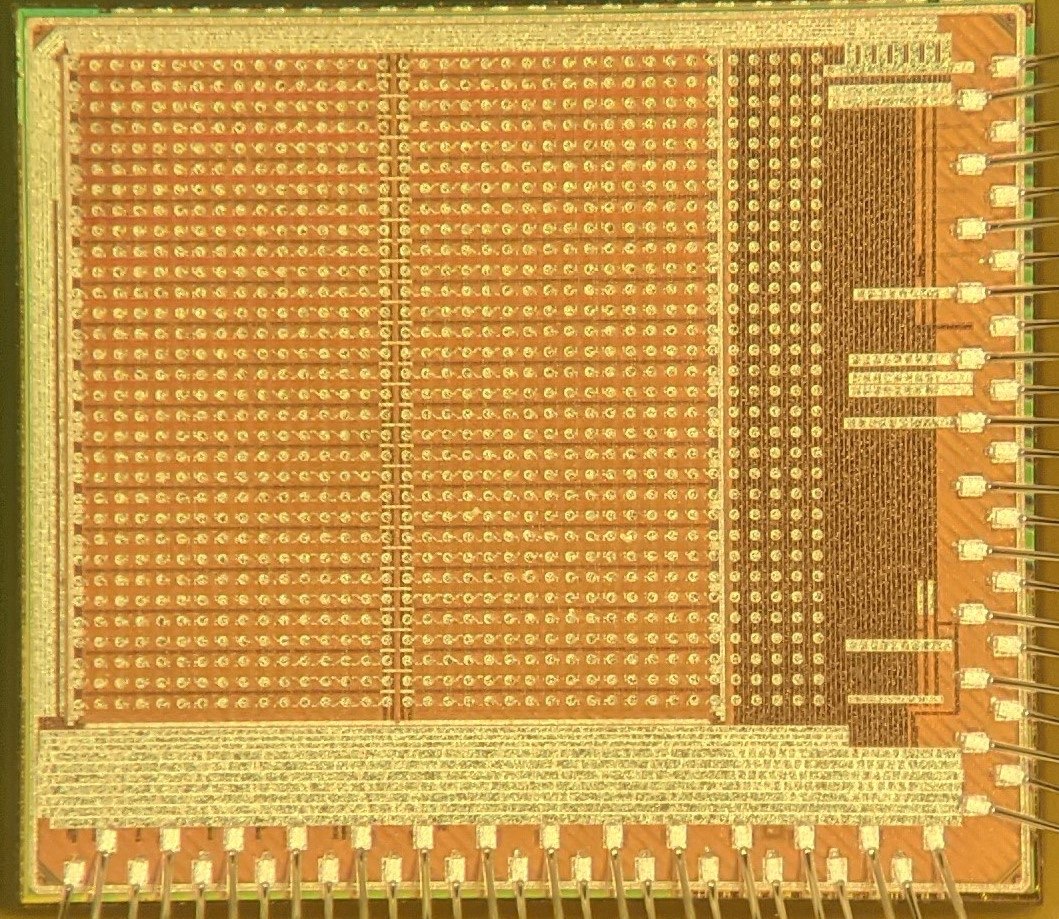}
        \caption{Photograph of the Timespot1 silicon die.}
    \label{fig:silicon_die}
\end{figure}

Figure~\ref{fig:silicon_die} shows a microscope photograph of the Timespot1 die, after wire-bonding on the test board.

\begin{figure}
    \centering
    \includegraphics[width=0.7\columnwidth]{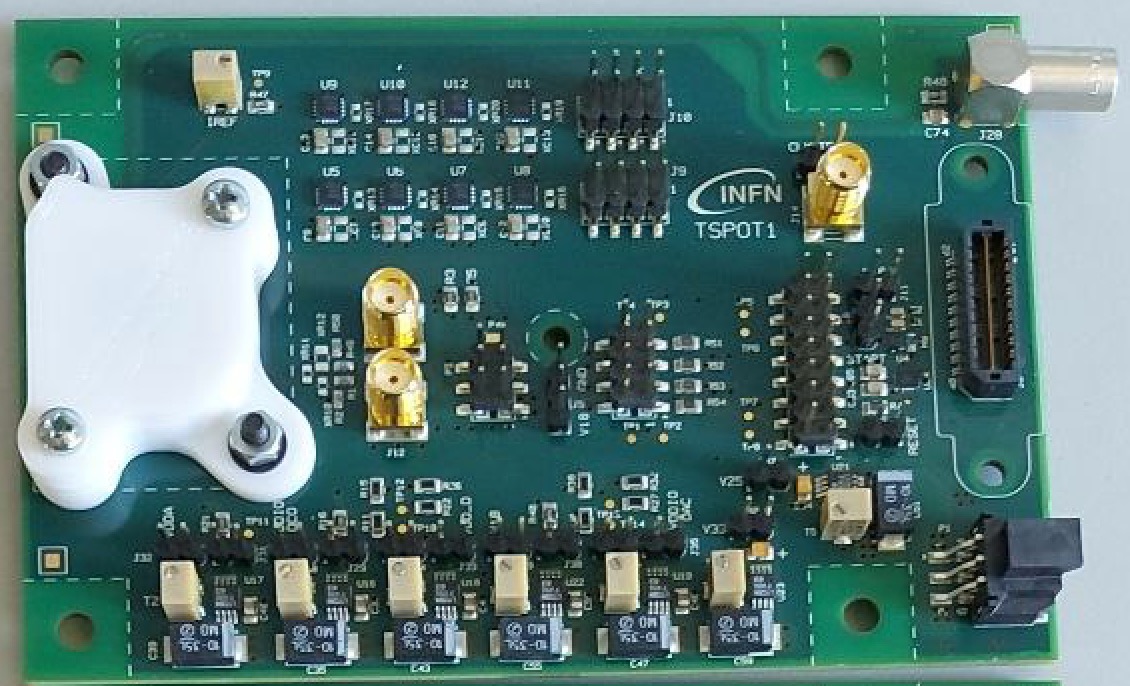}
        \caption{Photograph of the TSPOT1 PCB for the complete test of the Timespot1 ASIC. The PCB sizes \SI{80x120}{\milli\meter\squared}. The die and the wire-bonds are protected under the 3D-printed white box on the left.}
    \label{fig:pcb}
\end{figure}

Timespot1 is being tested concerning its operation and circuit performance on a dedicated Printed-Circuit Board, named TSPOT1 (Figure~\ref{fig:pcb}). The tests illustrated in this paper where performed on the front-end chip alone, before hybridization with the sensor. The TSPOT1 is conceived to have the maximum possible access to the ASIC I/O, allowing the possibility to input and then visualize the signals by means of a pattern generator and a high-performance logic analyzer or oscilloscope. The same board can be used with a high-performance \st{read-out}\textcolor{red}{readout} board, based on a Xilinx Kintex7 FPGA.

The test is performed using internal test pulses on individual channels.
The TDC is tested after performing the Vernier's DCO calibration. TA and ToT are verified using \st{7 and 32 points respectively} \textcolor{red}{the self-generated test pulses}, whereas the Front-End is tested using a test pulse generated internally with a fixed phase to the \SI{40}{\mega\hertz} clock.


\subsection{TDC Characterization}




The TDC is characterized by its TA resolution $\sigma_\mathrm{TDC}$, which is the most critical parameter for the timing performance.
\rnote{In particular $\sigma_\mathrm{TDC}$, represents the total TDC resolution which the main contributions comes from the quantization error, the measurement process uncertainty and the reference clock stability.  This last contribution has been evaluated using a loop-back configuration on the input clock driver. Its rms jitter has been evaluated to be around $\SI{10}{\pico\second}_\text{rms}$ at the chip input. However, the stability will degrade along the clock path. A punctual evaluation of the clock stability is not measurable, therefore this value must be taken as an indication of its lower bound.}
The ToT\st{, on the other hand,} has an \cor{resolution}{LSB} of about \SI{1}{\nano\second} which meets the performance required to adequately apply the ToT correction. 

\begin{figure}
    \centering
    \includegraphics[width=\columnwidth]{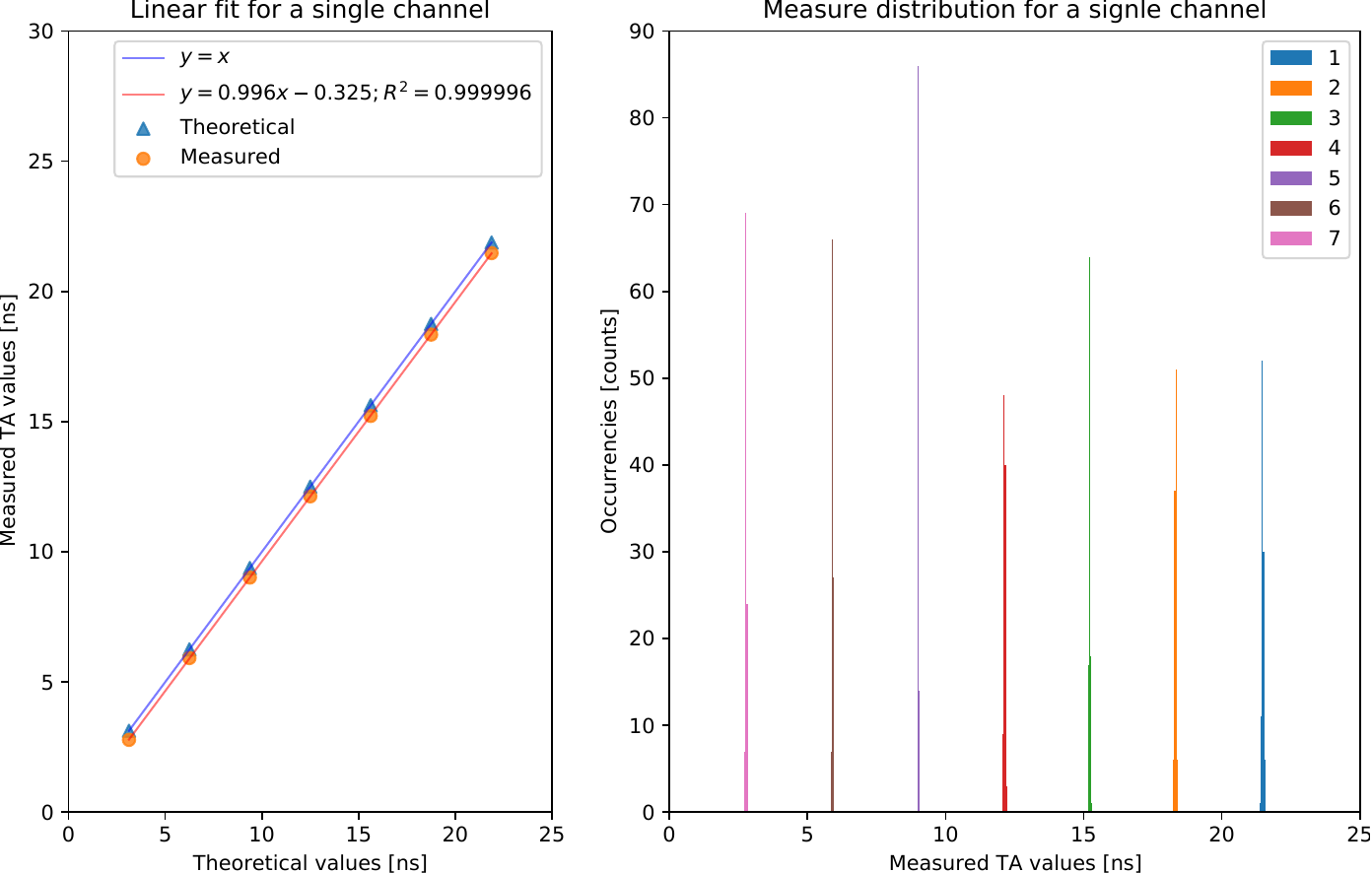}
    \caption{TA Linearity curve on the left and TA Histogram for 7 different injection phases on the right. On the linearity curve the errors bars, extracted from the phase histograms, are not visible at the plot scale.}
    \label{fig:TAhistoScan}
\end{figure}

\st{Each TDC is tested using the 7 points to verify the performance in all its dynamic range. The 7 test phases are distributed uniformly inside the range, in step of 3.125ns. A first outcome of such procedure is the TDC response linearity over the range.}  
\textcolor{red}{The first tests concerned the study of the linearity of the TDC response. This is done by pulsing the single TDC repeatedly with the 7 test points which are evenly spaced inside the dynamic range. At the end of the test, the fit line is calculated. The slope of this line, together with the standard deviation of the individual measurements with respect to the line and the coefficient of determination ($R^2$), provides us with a good indication of the linear response of the TDC and its INL. In Figure~\ref{fig:TAhistoScan} we can see a tipical TA scan result for a single channel. On the left we have the linearity response, with the equation of the line fitting the data, with the coefficient of determination $R^2$. In this plot we can also see a shift with respect to the theoretical straight line. This is expressed by the intercept value of the line which is non-zero. This is an expected phenomenon and due to a delay in the ignition system, which results in a systematic displacement of the points.}
Histograms of repeated measurements on each point are also \st{build} \textcolor{red}{built} in order to evaluate the \cor{precision for}{error associated to} each \cor{phase tested}{performed phase measurement}, as shown on the right plot of Figure~\ref{fig:TAhistoScan}.

\begin{figure}[t]
    \centering
    \includegraphics[width=\columnwidth]{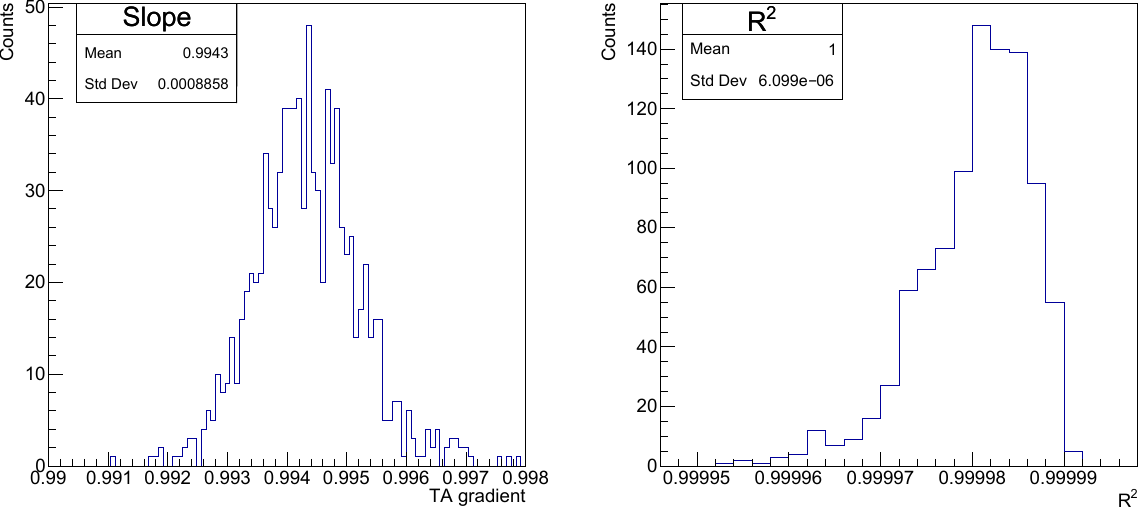}
    \caption{Results of linearity response of all channels, except broken ones. On the left we plot the slope distribution, while on the right the coefficient of determination ($R^2$) distribution.}
    \label{fig:TAslopeR2}
\end{figure}

The test is extended to all the \num{1024} channels, to evaluate the uniformity around the matrix, and also to \st{found all the} \textcolor{red}{identify} faulty channels. Figure~\ref{fig:TAslopeR2} shows the results of linearity response of all channels. From this plots were excluded the channels not working. On the left we have the resulting slope distribution, while on the right we can see the coefficient of determination ($R^2$) distribution.
\st{Some channels shows a sigma\_TDC = 0, that can be due to either a broken channel or a low statistics.}


\begin{figure*}
    \centering
    \includegraphics[width=\textwidth]{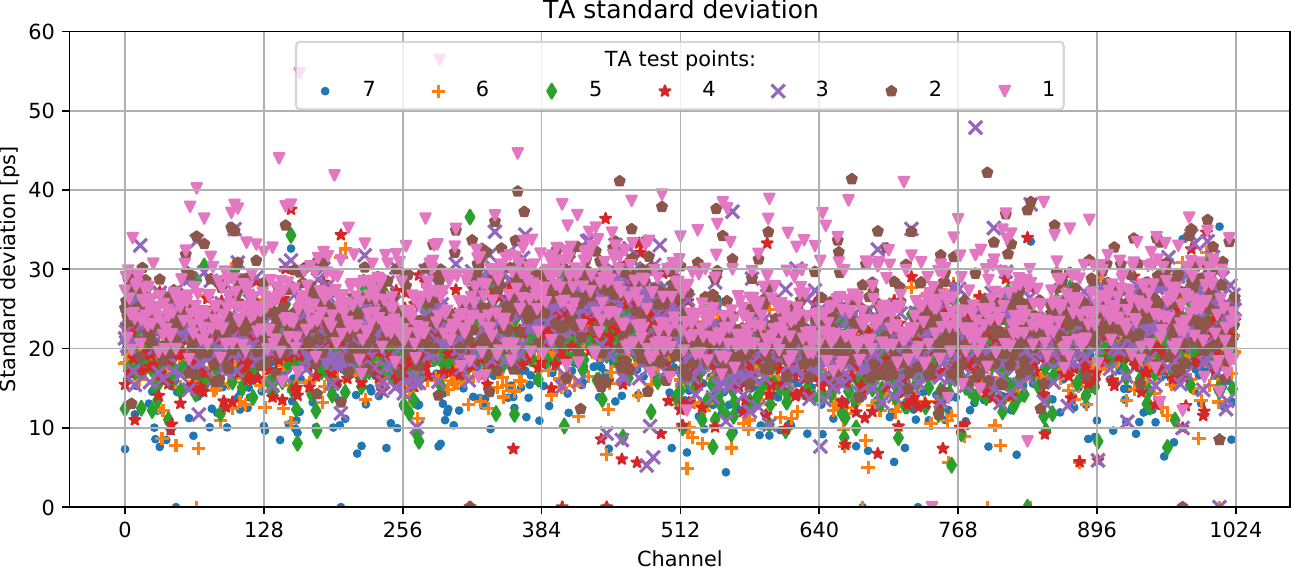}
    \caption{TA standard deviation across \num{1024} channels for the \num{7} input phases. Each point is computed from 100 measurements.}
    \label{fig:TAstdevAllCh}
\end{figure*}

Figure~\ref{fig:TAstdevAllCh} shows the scatter plot of the $\sigma_{\mathrm{TDC}}$ across all channel and phases, the phase and position dependencies can be observed in this plot. 
The plot shows a good uniformity across all matrix, with a small geometry effect on the right sub-matrix (after channel \num{512} as shown in Figure~\ref{fig:labeled_chip}). The plot also shows that TDC tends to perform slightly better for lower phase pulses. 

\begin{figure}
    \centering
    \includegraphics[width=0.59\columnwidth]{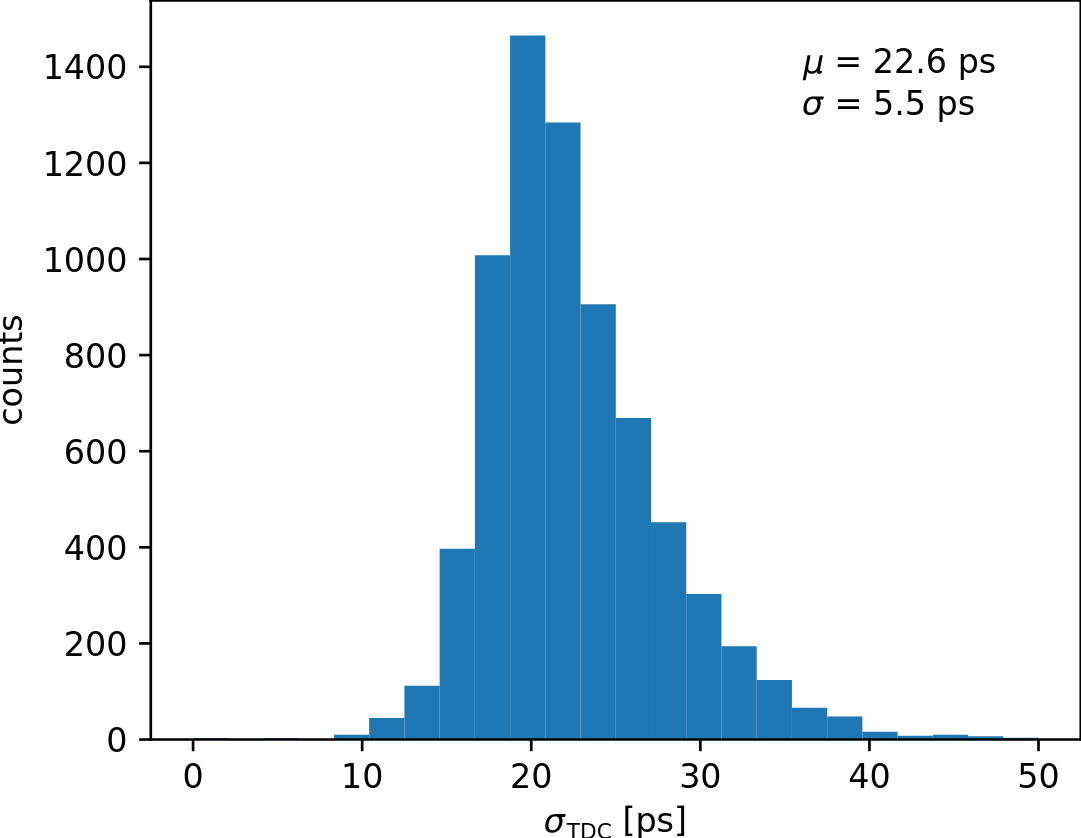}
        \caption{Histogram of the TA standard deviation measured on the TDC, across the \num{1024} channels and the seven test pulses.}
    \label{fig:TDC_histo} 
\end{figure}

Figure~\ref{fig:TDC_histo} shows the histogram of the distribution of the data of Figure~\ref{fig:TAstdevAllCh}. The average resolution is $\SI{22.6}{\pico\second}_\text{rms}$. Pixel position and signal phase account for a variation of $\SI{5.5}{\pico\second}_\text{rms}$ (standard deviation) from this mean value.

\subsection{Analog Front-End Characterization}

In the Timespot1 ASIC, the only method to measure the AFE output is by \st{leveraging} \textcolor{red}{taking advantage of} the channel TDC. Each AFE channel can be individually pulsed using its charge injection circuit. In response to this signal the AFE will produce a digital pulse which is then measured by the TDC.    Consequently, with a single measurement, it is only possible to measure the TA and ToT of a signal. In order to extract more information, the channel can be statistically investigated by repeatedly pulsing the same channel. In this way the \cor{time resolution}{output signal jitter} for a given condition can be evaluated as the standard deviation of TA measurement set ($\sigma_{\mathrm{AFE+TDC}}$).
 \st{However this contribution accounts for both the AFE and TDC time fluctuations.}  The AFE component $\sigma_{\mathrm{AFE}}$ can be evaluated by removing the previously measured TDC contribution $\sigma_{\mathrm{TDC}}$. This can be computed under the hypotheses that the two sources of variation are independent using:
\begin{equation} \label{eq:1}
	\sigma_{\mathrm{AFE}} =
	\begin{cases}
	  \sqrt{ \sigma_{\mathrm{AFE+TDC}}^2-\sigma_{\mathrm{TDC}}^2  } &  \text{ if } \sigma_{\mathrm{AFE+TDC}} \geq \sigma_{\mathrm{TDC}}\\
	  0   & \text{ if } \sigma_{\mathrm{AFE+TDC}} < \sigma_{\mathrm{TDC}}
	\end{cases}
 	\end{equation}
\noindent Contrary what is natural to expect, values of $\sigma_{\mathrm{AFE+TDC}}$ smaller then $\sigma_{\mathrm{TDC}}$ can be measured. This case can be explained as an effect of a too small statistic\textcolor{red}{s}: \cor{the measurement sets were too small to distinguish this difference.}{ given the small difference to be measured in these specific cases, a larger statistical set is required to distinguish it compared to the cases with a larger $\sigma_{\mathrm{AFE}}$.}

\begin{figure}[b]
    \centering
    \includegraphics[width=0.8\linewidth]{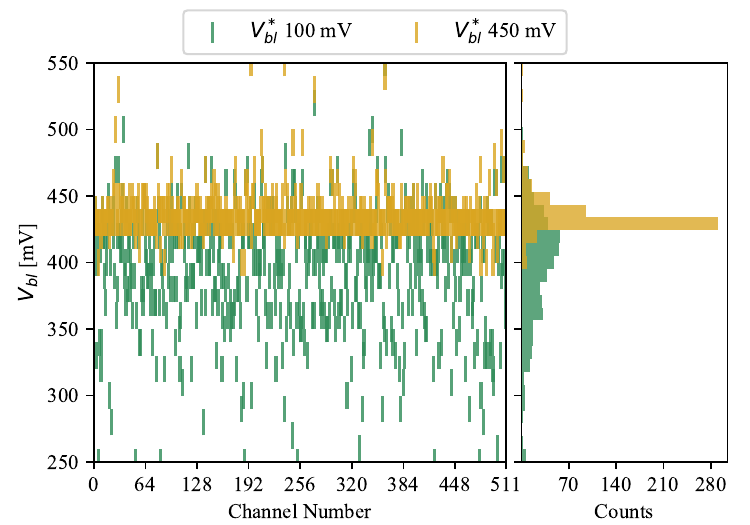}
    \caption{Measured baseline distribution $V_{bl}$ across 512 channels. The measurement was repeated for two desired baselines $V_{bl}^*$ \rnote{programmed with a dedicated DAC and set using offset compensation. The two set values are}: \SI{100}{mV} and \SI{450}{mV}.  It can be observed that the OC fails to adequately compensate the channels for low baseline values. Rising $V_{bl}^*$ to \SI{450}{mV} will move the channel population to the desired baseline.}
    \label{gr:bl}
\end{figure}

\rnote{In terms of the analog power consumption, the system can be finely configured by setting the bias currents on a per-block basis.  The values of the configuration steps have been verified by directly measuring the analog current intake using a source meter directly connected to the analog power net. This measurement are inline with what expected in simulation. The following measurements were performed with a power consumption of \SI{13}{\micro\watt} per channel. This value includes the static power consumption of all the analog circuits in the pixel and the periphery, averaged for the number of pixels.}

Moreover it is possible to reconstruct the CSA signal using a threshold-scan. During this analysis an issue with the OC was found: the OC has proven to be \st{inadequate to compensate baseline values to low voltages} \textcolor{red}{not able to compensate baseline fluctuations for low $V_{bl}$ voltages}. Figure~\ref{gr:bl} shows the baseline positions spread across \textcolor{red}{a} half matrix. \rnote{The actual baseline position $V_{bl}$ for each channel has been extracted from the threshold scan reconstruction, ant it has been defined as the level in which the signal starts forming.}  The analysis was performed with two desired baseline values $V_{bl}^*$\rnote{, this voltage is the one provided by a global DAC on the discriminator threshold terminal.} \cor{This problem is probably caused by an higher than expected voltage to be compensated. The OC time assigned by design is insufficient to discharge the capacitor to the desired level.}{This problem might be caused by an incorrect setting of the voltages on the discriminator threshold terminal.}

\st{The OC procedure correct} \textcolor{red}{The correct OC procedure} operation has already been verified in the previous prototype \cite{bib:testAFE0}. The only difference in the implementation of the discriminator between the two prototypes is the switching circuit.  In the current implementation the circuit has been changed to use p-type switches since they have proven to be more radiation resistant~\cite{bib:TID_28}. \cor{This change may have  increased the discriminator pre-OC DC voltage point.}{Simulation shows that the transistor forming the switch on the threshold terminal is not in conduction regime whenever a low voltage is applied. In this condition the node is biased by its source-bulk current to a value around \SI{300}{\milli\volt}, similar to the one found in the analysis of Figure \ref{gr:bl}.}

\begin{figure}[b]
    \centering
    \includegraphics[width=0.9\linewidth]{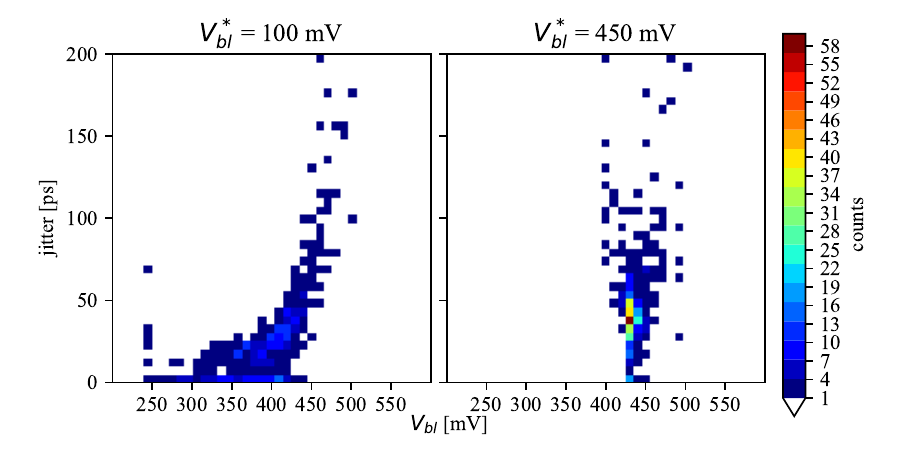}
    \caption{Correlation of $\sigma_{\mathrm{AFE}}$ to $V_{bl}$ for two $V_{bl}^*$ values.  \rnote{The plot is obtained by analysing threshold scans on different channels with an input charge of \SI{2}{\femto\coulomb}. The 2D histograms collects the jitter performance of the channels correlated with their baseline position.}  In the left plot the OC is not working properly. It can be observed a correlation between the AFE resolution and the baseline position. Measurements at low baseline values (on the left) suggest that the CSA intrinsic \cor{resolution}{jitter} is below the one of the TDC (\SI{20}{ps}). When the OC is working (right plot) the performance\st{s are} \textcolor{red}{is} in line with what is observed at the corresponding baseline in the left plot.  }
    \label{gr:vbl_jitter}
\end{figure}
\begin{figure}
    \centering
    \includegraphics[width=0.9\linewidth]{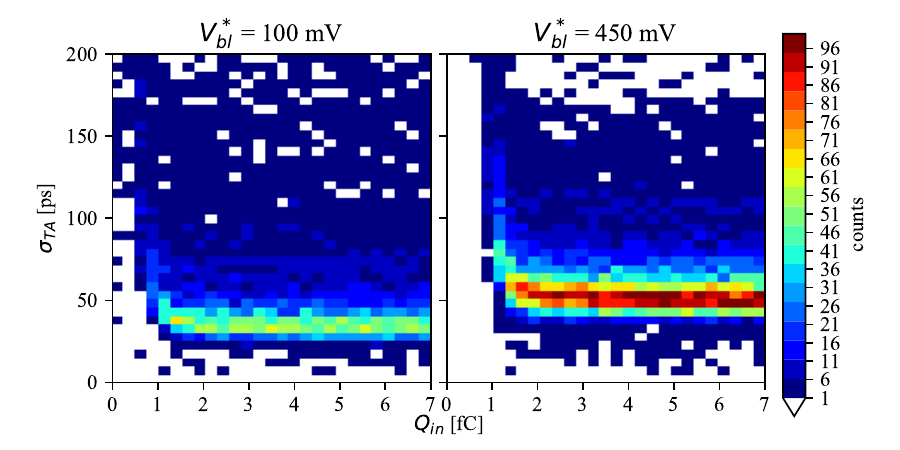}
    \caption{\rnote{2D histogram collecting} $\sigma_{\mathrm{AFE}}$ values as function of the input charge. As expected, the time resolution improves with larger input charges. Consistently with what is shown in Figure~\ref{gr:vbl_jitter}, the jitter performance is improved for low baselines, but the per-channel variation is improved when the OC is working properly. }
    \label{gr:sta_qin}
\end{figure}
\begin{figure}[]
    \centering
    \includegraphics[width=0.9\linewidth]{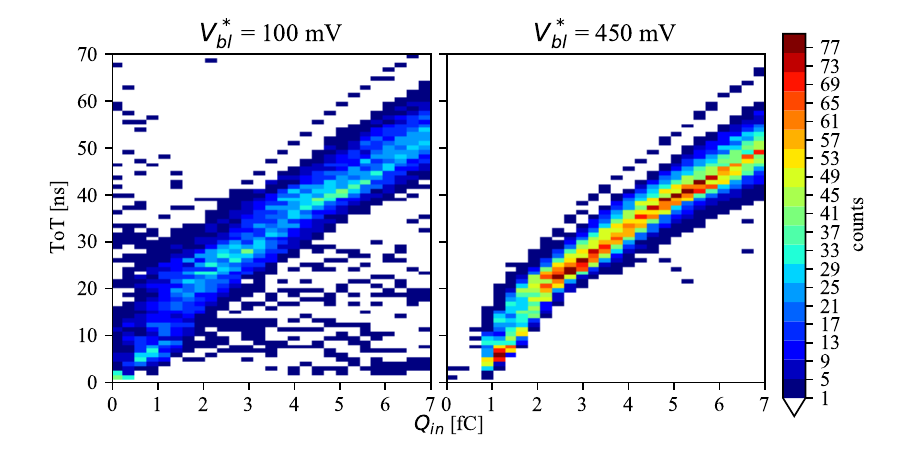}
    \caption{
    \rnote{2D histogram collecting} ToT as function of the input charge. The ToT shows a good linearity.
    Consistently with what it is shown in Figure \ref{gr:vbl_jitter}, channel variation improves when the OC is  working properly. }
    \label{gr:tot_qin}
\end{figure}

However the OC is still working properly for high $V_{bl}^*$ values. 
 The side effect of imposing a\st{n} high-baseline is related to the fact that the threshold value $V_{thr}$ must be set to an even higher value in order to measure signals with positive polarity.  In this range, the p-type input discriminator will not operate at its optimal common mode voltage, thus limiting its bandwidth.  This behavior is shown in Figure~\ref{gr:vbl_jitter} in which $\sigma_{\mathrm{AFE}}$ is correlated to the measured baseline $V_{bl}$. When the OC is not working properly a correlation between $V_{bl}$ and $\sigma_{\mathrm{AFE}}$ can be observed. Moreover, when the OC is set to high values, the compensated channels will feature\st{s} the same $\sigma_{\mathrm{AFE}}$ which was found with the corresponding baselines in the previous case. \cor{This two facts suggest that the discriminator core is band-width limited for high input voltages}{These two facts suggest that the discriminator core is band-width limited for the high DC operating points: in this condition the biasing transistor of the differential pair drains a lower current than the desired one. This behavior has been verified in simulation.}

In such conditions, the AFE time resolution has been evaluated for a \SI{2}{\femto\coulomb} signal when the OC is working properly.  This resulted in an average of $\SI{43}{\pico\second}_\text{rms}$ for $\sigma_{\mathrm{AFE}}$ with a power consumption around \SI{13}{\micro\watt} per channel.  However, if the hypothesis of the discriminator malfunction is true, the CSA \cor{is proven to}{should} be able to reach a resolution better than $\SI{20}{\pico\second}_\text{rms}$.

\begin{figure}
    \centering
    \includegraphics[width=0.9\linewidth]{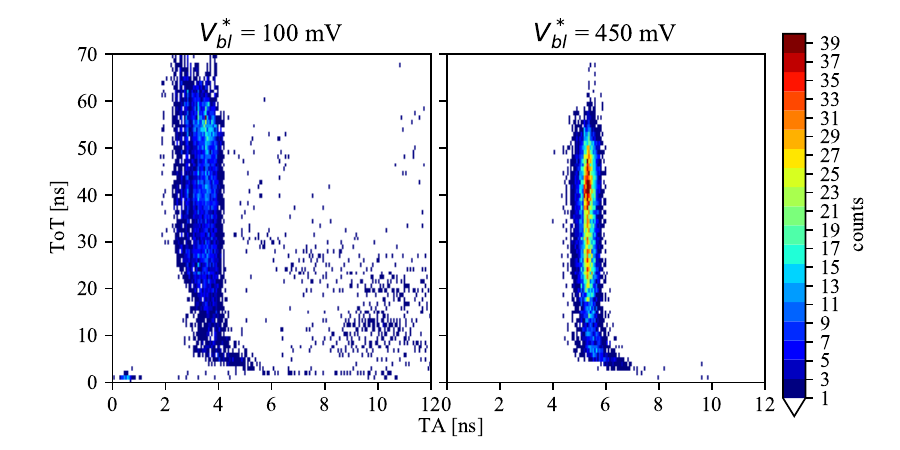}
    \caption{
    \rnote{2D histogram collecting} TA versus ToT correlation. This relation is used to perform a ToT-correction of the time-walk effect.
    Consistently with what is shown in Figure~\ref{gr:vbl_jitter}, channel variation improves when the OC is working properly. 
    }
    \label{gr:ta_vs_tot}
\end{figure}

Another important characterization of the AFE is its performance for different input charges.  Figure~\ref{gr:sta_qin} shows $\sigma_{\mathrm{AFE}}$ for different input charges.    Figure~\ref{gr:tot_qin} shows the ToT linearity.  Finally, Figure~\ref{gr:ta_vs_tot} reports the relation of the signal TA and its ToT.  This relation is useful to correct the time-walk effect. It can be observed in each case that the issue with the OC is present for low baselines.  As expected, the correct operation of the OC improves the per-channel signal variation. 

\section{Conclusions}

The characterization of the Timespot1 ASIC has been presented. Timespot1 is the first \SI{28}{\nano\meter} CMOS ASIC \st{ever} designed and fabricated for the readout of pixel matrices with high-resolution \st{time} \textcolor{red}{timing} capabilities per pixel. Still on a reduced size of \num{32x32} pixels, with a \SI{55}{\micro\meter} pitch, it implements the full set of functionalities needed for the purpose of 4D tracking.

The Timespot1 AFE is capable of an average time resolution of $\SI{43}{\pico\second}_\text{rms}$, with a measured \st{value} dispersion across the pixel matrix of about \SI{18}{\pico\second}. The TDC has an average time resolution of $\SI{22.6}{\pico\second}_\text{rms}$ and a \st{value} dispersion of \SI{5.5}{\pico\second}. 

The measured AFE performance is affected by an identified design bug, causing the Offset-Compensating procedure to be inadequate for proper offset compensation when the threshold is set to low values. Once this bug is corrected, the performance of the AFE should be capable of a resolution of the order \cor{or below}{of} $\SI{20}{\pico\second}_\text{rms}$ \rnote{or below}.
\st{On the other hand, the resolution of the TDC appears to be limited by the system clock stability and should be significantly improved after a more accurate clock distribution across the matrix area.} 
\rnote{On the other hand, given the TDC bin size of around \SI{50}{\pico\second}, the TDC quantization error of $\SI{14}{\pico\second}_\text{rms}$ approaches the supposed clock stability. It is reminded that is value is greater than $\SI{10}{\pico\second}_\text{rms}$, making it a limiting factor for the overall resolution. Therefore the resolution can be significantly improved after a more accurate clock distribution across the matrix area.}

The ASIC has been recently hybridized on 3D-trench silicon sensors~\cite{3D-TRENCH} featuring a pixel matrix with matched geometry. Using the same TSPOT1-PCB, the hybrid is being tested stand-alone in the laboratory, stimulated by an IR pulsed laser source. Finally, the TSPOT1-PCB, equipped with the hybrid, will be used as a layer in a tracking telescope in test-beams with MIPs.

In the meantime, the design of a new version, featuring a larger pixel matrix of \num{64x64} or \num{128x128} pixels and improved timing performance is near to start.
\rnote{Such a large area matrix will be based on the current matrix, which is intended to be easily duplicated and repeated. To overcome the typical problems of large area matrices related to power distribution, it is thought to use the Through Silicon Vias (TSV), in order to distribute the power more punctually, reducing the power drops. Also the use of TSV can improve the distribution of the clock reducing the jitter on the clock linked to its distribution inside the chip, thus allowing to increase the temporal resolution achievable.}

\bibliographystyle{JHEP}
\bibliography{bib.bib}

\providecommand{\href}[2]{#2}\begingroup\raggedright\begin{thebibliography}{10}

\bibitem{LHCB_TDR}
M.~van Beuzekom et~al., \emph{Vertex detector for {LHCb} {U}pgrade-{II}},  in
  \emph{JPS Conf. Proc.}, vol.~34, p.~010014, 2021.

\bibitem{TIMESPOT2018}
A.~Lai, \emph{A system approach towards future trackers at high luminosity
  colliders: the {TIMESPOT} project},  in \emph{IEEE Nuclear Science Symposium
  and Medical Imaging Conference Proceedings (NSS/MIC)}, pp.~1--3, 2018,
  \href{https://doi.org/10.1109/NSSMIC.2018.8824310}{DOI}.

\bibitem{LAI2020164491}
A.~Lai, L.~Anderlini, M.~Aresti, A.~Bizzeti, A.~Cardini, G.-F.~Dalla~Betta
  et~al., \emph{{First results of the TIMESPOT project on developments on fast
  sensors for future vertex detectors}},
  \href{https://doi.org/https://doi.org/10.1016/j.nima.2020.164491}{\emph{Nucl.
  Instrum. Meth. A} {\bfseries 981} (2020) 164491}.

\bibitem{3D-TRENCH}
L.~Anderlini, M.~Aresti, A.~Bizzeti, M.~Boscardin, A.~Cardini,
  G.-F.~Dalla~Betta et~al., \emph{{Intrinsic time resolution of 3D-trench
  silicon pixels for charged particle detection}},
  \href{https://doi.org/10.1088/1748-0221/15/09/p09029}{\emph{J.Instrum.}
  {\bfseries 15} (2020) P09029}.

\bibitem{3D-accurate}
D.~Brundu, A.~Cardini, A.~Contu, G.~Cossu, G.-F.~Dalla~Betta, M.~Garau et~al.,
  \emph{{Accurate modelling of 3D-trench silicon sensor with enhanced timing
  performance and comparison with test beam measurements}},
  \href{https://doi.org/10.1088/1748-0221/16/09/p09028}{\emph{J.Instrum.}
  {\bfseries 16} (2021) P09028}.

\bibitem{bib:TID_28}
C.-M.~Zhang, F.~Jazaeri, G.~Borghello, S.~Mattiazzo, A.~Baschirotto and C.~Enz,
  \emph{Bias dependence of total ionizing dose effects on 28-nm bulk
  {MOSFETs}},  in \emph{IEEE Nuclear Science Symposium and Medical Imaging
  Conference Proceedings (NSS/MIC)}, pp.~1--3, 2018,
  \href{https://doi.org/10.1109/NSSMIC.2018.8824379}{DOI}.

\bibitem{bib:65rad1}
M.~Krohn, B.~Bentele, D.~Christian, J.~Cumalat, G.~Deptuch, F.~Fahim et~al.,
  \emph{Radiation tolerance of 65 nm cmos transistors},
  \href{https://doi.org/10.1088/1748-0221/10/12/P12007}{\emph{Journal of
  Instrumentation} {\bfseries 10} (2015) P12007}.

\bibitem{bib:65rad2}
S.~Bonacini, P.~Valerio, R.~Avramidou, R.~Ballabriga, F.~Faccio, K.~Kloukinas
  et~al., \emph{Characterization of a commercial 65 nm cmos technology for slhc
  applications},
  \href{https://doi.org/10.1088/1748-0221/7/01/P01015}{\emph{Journal of
  Instrumentation} {\bfseries 7} (2012) P01015}.

\bibitem{bib:16rad}
T.~Ma, S.~Bonaldo, S.~Mattiazzo, A.~Baschirotto, C.~Enz, A.~Paccagnella et~al.,
  \emph{Tid degradation mechanisms in 16-nm bulk finfets irradiated to
  ultrahigh doses}, \href{https://doi.org/10.1109/TNS.2021.3076977}{\emph{IEEE
  Transactions on Nuclear Science} {\bfseries 68} (2021) 1571}.

\bibitem{bib:cms}
{The CMS Collaboration}, \emph{Technical proposal for a {MIP} timing detector
  in the {CMS} experiment {P}hase 2 upgrade},  Tech. Rep. CERN, Geneva,
  Switzerland (Nov., 2017).

\bibitem{bib:atlas}
{The ATLAS Collaboration}, \emph{High-granularity timing detector for the
  {ATLAS} {P}hase-{II} upgrade},  Technical Proposal CERN, Geneva, Switzerland
  (July, 2018).

\bibitem{bib:FTDRLHCbU2}
{The LHCb Collaboration}, \emph{Framework {TDR} for the {LHCb} upgrade {II}:
  Opportunities in flavour physics, and beyond, in the {HL-LHC} era},  Tech.
  Rep. CERN, Geneva, Switzerland (Feb., 2022).

\bibitem{bib:na62}
{The KOTO, LHCb, and NA62/KLEVER Collaborations, and the US Kaon Interest
  Group}, \emph{Searches for new physics with high-intensity kaon beams},
  Apr., 2022.
\newblock 10.48550/ARXIV.2204.13394.

\bibitem{bib:tofhir2x}
T.S.~Niknejad, \emph{Results with the {TOFHIR2X} revision of the front-end
  {ASIC} of the {CMS} {MTD} barrel timing layer},  Tech. Rep. CERN, Geneva,
  Switzerland (Dec., 2021).

\bibitem{bib:lhcbcooling}
O.A.~{de Aguiar Francisco}, W.~Byczynski, K.~Akiba, C.~Bertella, A.~Bitadze,
  M.~Brock et~al., \emph{Microchannel cooling for the {LHCb} {VELO} upgrade
  {I}},
  \href{https://doi.org/https://doi.org/10.1016/j.nima.2022.166874}{\emph{Nucl.
  Instrum. Meth. A} {\bfseries 1039} (2022) 166874}.

\bibitem{bib:MonolithicRadHard}
C.~Riegel, M.~Backhaus, J.V.~Hoorne, T.~Kugathasan, L.~Musa, H.~Pernegger
  et~al., \emph{Radiation hardness and timing studies of a monolithic
  {T}ower{J}azz pixel design for the new {ATLAS} {I}nner {T}racker},
  \href{https://doi.org/10.1088/1748-0221/12/01/C01015}{\emph{Journal of
  Instrumentation} {\bfseries 12} (2017) }.

\bibitem{bib:3d}
S.I.~Parker, C.J.~Kenney and J.~Segal, \emph{{3D} -- {A} proposed new
  architecture for solid-state radiation detectors},
  \href{https://doi.org/10.1016/S0168-9002(97)00694-3}{\emph{Nucl. Instrum.
  Meth. A} {\bfseries 395} (1997) 328}.

\bibitem{bib:tpix_42022}
X.~Llopart, J.~Alozy, R.~Ballabriga, M.~Campbell, R.~Casanova, V.~Gromov
  et~al., \emph{{Timepix4, a large area pixel detector readout chip which can
  be tiled on 4 sides providing sub-200 ps timestamp binning}},
  \href{https://doi.org/10.1088/1748-0221/17/01/c01044}{\emph{{J. Instrum.}}
  {\bfseries 17} (2022) C01044}.

\bibitem{bib:fastpix}
J.~Braach, E.~Buschmann, D.~Dannheim, K.~Dort, T.~Kugathasan, M.~Munker et~al.,
  \emph{Performance of the {FASTPIX} sub-nanosecond {CMOS} pixel sensor
  demonstrator},
  \href{https://doi.org/10.3390/instruments6010013}{\emph{Instruments}
  {\bfseries 6} (2022) }.

\bibitem{bib:fast2}
``{FAST2: a new family of front-end ASICs to read out thin Ultra-Fast Silicon
  detectors achieving picosecond time resolution}.''
  \url{https://indico.cern.ch/event/1019078/contributions/4443951/}, 2021.

\bibitem{bib:fastic}
``{FastIC: A Fast Integrated Circuit for the Readout of High Performance
  Detectors}.''
  \url{https://indico.cern.ch/event/1019078/contributions/4443966/}, 2021.

\bibitem{bib:diamasic}
A.~Ghimouz, F.E.~Rarbi and O.~Rossetto, \emph{{DIAMASIC}: A multichannel
  front-end electronics for high-accuracy time measurements for diamond
  detectors},  2021.
\newblock 10.48550/ARXIV.2110.12440.

\bibitem{bib:etroc1}
H.~Sun, D.~Gong, W.~Zhang, C.~Edwards, G.~Huang, X.~Huang et~al.,
  \emph{Characterization of the {CMS} endcap timing layer readout chip
  prototype with charge injection},
  \href{https://doi.org/10.1088/1748-0221/16/06/p06038}{\emph{Journal of
  Instrumentation} {\bfseries 16} (2021) P06038}.

\bibitem{bib:altiroc1}
C.~Agapopoulou, P.~Dinaucourt, A.~Dragone, D.~Gong, C.~de~La~Taille, N.~Makovec
  et~al., \emph{{ALTIROC} 1, a 25 ps time resolution {ASIC} for the {ATLAS}
  high granularity timing detector},  in \emph{IEEE Nuclear Science Symposium
  and Medical Imaging Conference (NSS/MIC)}, pp.~1--4, 2020,
  \href{https://doi.org/10.1109/NSS/MIC42677.2020.9507972}{DOI}.

\bibitem{bib:fcfd0}
``{Precision timing ASIC for LGAD sensors based on a Constant Fraction
  Discriminator – FCFD0 }.''
  \url{https://indico.cern.ch/event/1019078/contributions/4443948/}, 2021.

\bibitem{TDCpix}
G.~Aglieri~Rinella, S.~Bonacini, P.~Jarron, J.~Kaplon, A.~Kluge, M.~Morel
  et~al., \emph{{The TDCpix ASIC: High rate readout of hybrid pixels with
  Timing Resolution Better than 200 ps}},  in \emph{2013 IEEE Nuclear Science
  Symposium and Medical Imaging Conference (2013 NSS/MIC)}, pp.~1--4, 2013,
  \href{https://doi.org/10.1109/NSSMIC.2013.6829432}{DOI}.

\bibitem{NXPI2C7}
``{I2C-bus specification and user manual}.''
  \url{https://www.nxp.com/docs/en/user-guide/UM10204.pdf}, Oct., 2021.

\bibitem{bib:phdlpiccolo}
L.~Piccolo, \emph{An Analog Pixel Front-End for High Granularity Space-Time
  Measurements}, Ph.D. thesis, Politecnico di Torino Electrical, Electronics
  and Communications Engineering, 2022.

\bibitem{bib:krum}
F.~Krummenacher, \emph{Pixel detectors with local intelligence: an {IC}
  designer point of view}, {\emph{Nucl. Instrum. Meth. A} {\bfseries 305}
  (1991) 527 }.

\bibitem{ocmes}
L.~Piccolo, \emph{First measurements on a discrete-time front-end in 28-nm cmos
  technology for timing pixel detectors},  in \emph{2019 IEEE Nuclear Science
  Symposium and Medical Imaging Conference (NSS/MIC)}, pp.~1--4, 2019,
  \href{https://doi.org/10.1109/NSS/MIC42101.2019.9059904}{DOI}.

\bibitem{bib:VernTech}
R.G.~Baron, \emph{The vernier time-measuring technique},
  \href{https://doi.org/https://doi.org/10.1109/JRPROC.1957.278252}{\emph{Proc.
  of the IRE} {\bfseries 45} (1957) 527 }.

\bibitem{FISCHER2001499}
P.~Fischer, \emph{First implementation of the {MEPHISTO} binary readout
  architecture for strip detectors},
  \href{https://doi.org/https://doi.org/10.1016/S0168-9002(00)01283-3}{\emph{Nucl.
  Instrum. Meth. A} {\bfseries 461} (2001) 499}.

\bibitem{bib:testAFE0}
L.~Piccolo, \emph{First measurements on a discrete-time front-end in 28-nm
  {CMOS} technology for timing pixel detectors},  in \emph{IEEE Nuclear Science
  Symposium and Medical Imaging Conference (NSS/MIC)}, pp.~1--4, 2019,
  \href{https://doi.org/10.1109/NSS/MIC42101.2019.9059904}{DOI}.

\end{thebibliography}\endgroup

\end{document}